\documentclass[preprintnumbers,amsmath,amssymb,floatfix,14pt,prd,ocolumn,showpacs,
superscriptaddress,nofootinbib]{revtex4}
\usepackage[colorlinks=true, pdfstartview=FitV, linkcolor=blue, citecolor=red, urlcolor=magenta, breaklinks=true]{hyperref}
\usepackage[dvips]{graphicx}

\usepackage{epsfig}
\usepackage{graphicx}
\usepackage{indentfirst}
\usepackage{amsmath}
\usepackage{amsfonts}
\usepackage{amssymb}
\usepackage{array}
\begin{document}
\title{ $f(T,\mathcal{T})$  Cosmological Models in Phase Space}
\author{M. G. Ganiou}
\email{moussiliou_ganiou @yahoo.fr}
\affiliation{Institut de Math\'ematiques et de Sciences Physiques
(IMSP), Universit\'e de Porto-Novo, 01 BP 613 Porto-Novo, B\'enin}
\author{Ines G. Salako}
\email{inessalako@gmail.com}
\affiliation{Institut de Math\'ematiques et de Sciences Physiques
(IMSP), Universit\'e de Porto-Novo, 01 BP 613 Porto-Novo, B\'enin}
\affiliation{Ecole de Machinisme Agricole et de Construction M\'ecanique (EMACoM),  Universit\'e d'Agriculture de K\'etou, BP 13 K\'etou,  B\'enin}
\author{ M. J. S. Houndjo}
\email{ sthoundjo@yahoo.fr}
\affiliation{Institut de Math\'ematiques et de Sciences Physiques
(IMSP), Universit\'e de Porto-Novo, 01 BP 613 Porto-Novo, B\'enin}
\affiliation{Facult\'e des Sciences et Techniques de Natitingou (FAST), Universit\'e de Natitingou, BP 72 Natitingou, B\'enin}
\author{J. Tossa }
\email{joel.tossa@imsp-uac.org}
\affiliation{Institut de Math\'ematiques et de Sciences Physiques
(IMSP), Universit\'e de Porto-Novo, 01 BP 613 Porto-Novo, B\'enin}

\begin{abstract}
\hspace{0,2cm} 
In this paper we explore $f(T, \mathcal{T})$, where  $T$ and $\mathcal{T}$ denote the torsion scalar and the trace of the energy-momentum tensor 
respectively. We impose the covariant conservation to the energy-momentum tensor and obtain a cosmological $f(T, \mathcal{T})$ respectively. 
We impose the covariant conservation to the energy-momentum tensor and obtain a cosmological $f(T, \mathcal{T})$ model.
Then, we study the stability of the obtained model for power-law and de Sitter solutions and our result show that the model can be  stable for
some values of the input parameters, for both power-law and de Sitter solutions.
\end{abstract}
\maketitle
\pretolerance10000

\tableofcontents
\section{Introduction}\label{sec1}
Nowadays the current acceleration of the expansion of the universe is widely confirmed by several independent cosmological observational
data as Cosmic Microwave Background Radiation (CMBR) \cite{Spergel} and the Sloan Digital Sky Survey (SDSS) \cite{Adelman}.
This stage of the universe is explained in the literature through two approaches.
The first assumes that the universes if filled by an exotic ith negative pressure, named dark energy known as the responsible of this acceleration of 
the universe. The second approach, instead of assuming an exotic component, consists to modify the GR by changing the usual Einstein-Hilbert 
gravitational term, and various theories have been developed in this way and based on the Levi-Civita's connections, as 
( $f(R)$, $f(R,\mathcal{T})$ \cite{ma1}-\cite{ma6}, $f(G)$) \cite{mj1}-\cite{mj5} where $R$ denotes the curvature scalar,
$\mathcal{T}$ the trace of the energy-momentum tensor  and $G$ the Gauss-Bonnet invariant defined
by $G=R^2-4R_{\mu\nu}R^{\mu\nu} + R_{\mu\nu \
lambda\sigma}R^{\mu\nu\lambda\sigma}$.  There exists another type theory based the weitzenbock's connections, equivalent to GR,
called Tele-parallel Theory $(TT)$. This theory has been introduced by Ferraro {\it et al} \cite{st1} where 
they explained the UV modifications to the $TT$ and also the inflation. After this, Ferraro and Bengochea
 \cite{st1} have consider the same model to describe the dark energy.  other works can be found in   \cite{st2}- \cite{st41}.
In the same, modified versions of this theory have been developed and the one to which 
we are interested in this paper the $f(T, \mathcal{T})$, $T$ and $\mathcal{T}$ being the torsion scalar and the trace of
the energy-momentum tensor, respectively. 
 Specifically, this theory can be view as an
 homologue to $f(,\mathcal{T})$.
 Beside several works developed within 
 $f(T,\mathcal{T})$ \cite{sala1}- \cite{sala4}, we can note the one undertaken by Alvarenga and collaborators \cite{papierdiego}, 
 where they search for the model for which the covariant conservation of the energy-momentum tensor is realized. In that paper  
 they investigate the dynamics of scalar perturbations about the obtained model and focused  they attention to the sub-Hubble
 modes and show that  through the quasi-static approximation the result are very different from the ones derived in the frame
 of concordance $\Lambda CDM$, constraining the validity of this kind of model. \par
 In this paper we are interested to the coincidence cosmic problem and search for the $f(T,\mathcal{T})$ 
 model according to what the tress tensor is conserved.
 In order to obtain a consistent model, we explore the dynamics and stability about the obtained model. 
To reach our goal, we assume that it possible to have anti-gravity interactions between  the dark energy and the matter because of their unknown nature. We also introduce arbitrarily the terms of interaction between these components because we have not sure of the form of interaction between them. We realize a system of  three dynamic equations which take into account the dark energy, the dark matter and the ordinary matter. Consequently, we  reconstruct four models  and we show that the dynamic equations have two possible attractive solutions namely the phase dominated by the dark matter and that dominated by the dark energy. During this investigation, we have realized that some dynamic systems are unstable;  meaning that  a model provides that everything disappear in the Universe and leads to an Universe  more and more poor in energy. Other models show that the Universe should be filled by dark energy.  Another important feature emerging from this work is the stability study of $\Lambda$CDM model under 
consideration 
   by considering two interesting cosmological solutions i.e the power-law and the de Sitter solutions. We have analyzed  the constrains on the input parameters and as results, we have found that the stability is always realized.\par 
	This paper is organized as follows: In Sec \ref{sec2} we have reconstructed  a model by vanishing  the covariant derived of energy-momentum tensor. The stability of the obtained critical points of the dynamic systems has been explored in Sec \ref{sec3} and the perturbation functions have been determined within the model under consideration in Sec  \ref{sec5}. The Sec  \ref{sec4} is devoted  to  cosmological dynamic study of the considered model. Finally, we have ended our investigation by a conclusion in Sec \ref{sec6}.

 \section{Generality on $f(T,\mathcal{T})$  gravity within  FLRW Cosmology}\label{sec2}
 
 The modified theories of Tele-Parallel gravity  are those for which the scalar torsion of Tele-Parallel action is substituted by an arbitrarily function of this latter. As it is done in Tele-Parallel, the modified versions of  this theory are also described by the orthonormal tetrads which components are defined on the tangent space of each point of the manifold. The line element is written as
\begin{eqnarray}
ds^2=g_{\mu\nu}dx^\mu dx^\nu=\eta_{ij}\theta^i\theta^j\,, 
\end{eqnarray}
with the following definitions
\begin{eqnarray}
d^\mu=e_{i}^{\;\;\mu}\theta^{i}; \,\quad \theta^{i}=e^{i}_{\;\;\mu}dx^{\mu}.
\end{eqnarray}
 Note that  $\eta_{ij}=diag(1,-1,-1,-1)$ is the Minkowskian metric  and the $\{e^{i}_{\;\mu}\}$ are the components of the tetrad which satisfy the following identity:
\begin{eqnarray}
e^{\;\;\mu}_{i}e^{i}_{\;\;\nu}=\delta^{\mu}_{\nu},\quad e^{\;\;i}_{\mu}e^{\mu}_{\;\;j}=\delta^{i}_{j}.
\end{eqnarray}
In General Relativity, one use the following Levi-Civita's connection which  preserves the curvature whereas the torsion vanishes
\begin{equation}
\overset{\circ }{\Gamma }{}_{\;\;\mu \nu }^{\rho } =
\frac{1}{2}g^{\rho \sigma }\left(
\partial _{\nu} g_{\sigma \mu}+\partial _{\mu}g_{\sigma \nu}-\partial _{\sigma}g_{\mu \nu}\right)\;,
\end{equation}
But in the Tele-Parallel theory and its modified version, one keeps the scalar torsion by using Weizenbock's connection defined as:
\begin{eqnarray}
\Gamma^{\lambda}_{\mu\nu}=e^{\;\;\lambda}_{i}\partial_{\mu}e^{i}_{\;\;\nu}=-e^{i}_{\;\;\mu}\partial_\nu e_{i}^{\;\;\lambda}.
\end{eqnarray}

From this connection, one obtains the geometric objects. The first is the torsion defined by

\begin{eqnarray}
T^{\lambda}_{\;\;\;\mu\nu}= \Gamma^{\lambda}_{\mu\nu}-\Gamma^{\lambda}_{\nu\mu},
\end{eqnarray}
from which we define the contorsion as
\begin{equation}\label{K}
K_{\;\;\mu \nu }^{\lambda} \equiv \widetilde{\Gamma} _{\;\mu \nu }^{\lambda }
-\overset{\circ}{\Gamma }{}_{\;\mu \nu }^{\lambda}=\frac{1}{2}(T_{\mu }{}^{\lambda}{}_{\nu }
+ T_{\nu}{}^{\lambda }{}_{\mu }-T_{\;\;\mu \nu }^{\lambda})\;,
\end{equation}
Where the expression $\overset{\circ }{\Gamma }{}_{\;\;\mu \nu }^{\lambda}$ designs the above defined connection. Then we can write
\begin{eqnarray}
K^{\mu\nu}_{\;\;\;\;\lambda}=-\frac{1}{2}\left(T^{\mu\nu}_{\;\;\;\lambda}-T^{\nu\mu}_{\;\;\;\;\lambda}+T^{\;\;\;\nu\mu}_{\lambda}\right)\,\,.
\end{eqnarray}
The two previous geometric objects (the torsion and the contorsion) are used to define another tensor by 
\begin{eqnarray}
S_{\lambda}^{\;\;\mu\nu}=\frac{1}{2}\left(K^{\mu\nu}_{\;\;\;\;\lambda}+
\delta^{\mu}_{\lambda}T^{\alpha\nu}_{\;\;\;\;\alpha}-\delta^{\nu}_{\lambda}T^{\alpha\mu}_{\;\;\;\;\alpha}\right)\label{S}
\end{eqnarray}
From the fact that we are talking about the  modified versions of Tele-Parallel gravity, one use a general algebraic function of scalar torsion instead 
   the scalar torsion only as it is done in the initial theory. So, the new action is written as    
 
\begin{eqnarray}
 S= \int e \left[\frac{T+f(T,\mathcal{T})}{2\kappa^2} +\mathcal{L}_{m} \right]d^{4}x   \label{eq9}
\end{eqnarray}
where $\kappa^{2} = 8 \pi G $ is the usual constant coupling to Newton gravitational constant. Varying the action with respect to the tetrad, one obtains the equations of motion as \cite{sala1}- \cite{sala4} : 
\begin{eqnarray}\label{lagran1}
&&[\partial_\xi(ee^\rho_a
S^{\;\;\sigma\xi}_\rho)-ee^\lambda_a S^{\rho\xi\sigma} T_{\rho\xi\lambda}](1+f_T) 
+ e e^\rho_a(\partial_\xi T)S^{\;\;\sigma\xi}_\rho f_{TT} +\frac{1}{4} e e^\sigma_a (T) \nonumber \\
&&  =- \frac{1}{4} e e^\sigma_a \Big( f(\mathcal{T})\Big)  -e e^\rho_a(\partial_\xi \mathcal{T})S^{\;\;\sigma\xi}_\rho f_{T\mathcal{T}}  +
f_{\mathcal{T}}\;\Big(\frac{e\,\Theta^\sigma_{\;\;a}  
+ e e^\sigma_a \;p }{2}\Big) + \frac{\kappa^{2}}{2} e\,\Theta^\sigma_{\;\;a} \;,
\end{eqnarray}
with 
$f_{\mathcal{T}} = \partial f/\partial \mathcal{T} $,   $f_{T} = \partial f/\partial T$, $ f_{T\mathcal{T}} 
= \partial^{2}f/\partial T\partial \mathcal{T}$,
$f_{TT}  = \partial^{2}f/\partial T^{2}$ et  $\Theta^\sigma_{\;\;a}$ is the energy-momentum tensor of matter field.
 By using some transformations, we can establish the following relations:
 \begin{eqnarray}\label{nablaS'}
e^a_\nu e^{-1}\partial_\xi(ee^\rho_a
S^{\;\;\sigma\xi}_\rho)-S^{\rho\xi\sigma}T_{\rho\xi\nu} = -\nabla^\xi S_{\nu\xi}^{\;\;\;\;\sigma}-S^{\xi\rho\sigma}K_{\rho\xi\nu}\;,
\end{eqnarray}
\begin{equation}\label{eqdivs'}
G_{\mu\nu}-\frac{1}{2}\,g_{\mu\nu}\,T
=-\nabla^\rho S_{\nu\rho\mu}-S^{\sigma\rho}_{\;\;\;\;\mu}K_{\rho\sigma\nu}\;,
\end{equation}
By the end, from the combination of equations Eq.~(\ref{nablaS'}) and Eq.~(\ref{eqdivs'}, the field equations Eq.~(\ref{lagran1}) can be written as: 
\begin{equation} 
A_{\mu\nu}(1+ f_T) +\frac{1}{4}g_{\mu\nu}\;T =B_{\mu\nu}^{eff} \label{motion11}
\end{equation}
 where
\begin{eqnarray}\label{motion1add} 
&&A_{\mu \nu }=g_{\sigma\mu}e^a_\nu[e^{-1}\partial_\xi(ee^\rho_a
S^{\;\;\sigma\xi}_\rho)-e^\lambda_a S^{\rho\xi\sigma} T_{\rho\xi\lambda}]\\ \nonumber
&&\qquad=-\nabla^\sigma S_{\nu\sigma\mu }-S_{\;\;\;\;\mu }^{\rho\lambda }K_{\lambda \rho \nu }
=G_{\mu \nu }-\frac{1}{2}g_{\mu \nu }T, \; \\ \nonumber
&&B_{\mu\nu}^{eff} =  S^{\rho}_{\;\;\;\mu\nu}\; f_{T\mathcal{T}}\; \partial_{\rho} \mathcal{T}  -
S^{\rho}_{\;\;\;\mu\nu}\;f_{TT}\; \partial_{\rho} T  
  - \frac{1}{4} g_{\mu \nu }f +
f_{\mathcal{T}}\;\Big(\frac{\Theta_{\mu \nu }  
+ g_{\mu \nu } \;p }{2}\Big) + \frac{\kappa^{2}}{2} \,\Theta_{\mu \nu } \;,
\end{eqnarray}
So  the relation Eq.~(\ref{motion11}) can take the following form:
\begin{equation}
(1+ f_T)\,G_{\mu\nu}=T_{\mu\nu}^{eff} \label{motion12}
\end{equation}
 where
\begin{eqnarray}\label{motion1add'} 
T_{\mu\nu}^{eff} =S^{\rho}_{\;\;\;\mu\nu}\; f_{T\mathcal{T}}\; \partial_{\rho} \mathcal{T}  -
S^{\rho}_{\;\;\;\mu\nu}\;f_{TT}\; \partial_{\rho} T  - \frac{1}{4} g_{\mu \nu } \Big(T+ f \Big) +\frac{T\,g_{\mu\nu}\,f_T}{2}+
f_{\mathcal{T}}\;\Big(\frac{\Theta_{\mu \nu }  
+ g_{\mu \nu } \;p }{2}\Big) + \frac{\kappa^{2}}{2} \,\Theta_{\mu \nu } \;.
\end{eqnarray}
 \section{Reconstructing of model}\label{sec3}
In this section, we are interested to $  T+ Q\,\mathcal{T}^{\mathcal{N}} $  models which can reproduce the different features of $\Lambda CDM$.\par 
 In order to point out  the expression of the covariant energy-momentum tensor from which  one hopes extract a algebraic function, we take the  covariant derivative of (\ref{motion12}) which leads to:
\begin{eqnarray}
 \nabla^\mu \Big[(1+ f_T)\, G_{\mu\nu}\Big]& =&  \nabla^\mu T_{\mu\nu}^{eff}  \cr
 &=&   \nabla^\mu  \Bigg[\frac{T\,g_{\mu\nu}\,f_T}{2}+ S^{\rho}_{\;\;\;\mu\nu}\; f_{T\mathcal{T}}\; \partial_{\rho} \mathcal{T}  -
S^{\rho}_{\;\;\;\mu\nu}\;f_{TT}\; \partial_{\rho} T \nonumber\\&-& \frac{1}{4} g_{\mu \nu } \Big(T+ f \Big) +
f_{\mathcal{T}}\;\Big(\frac{\Theta_{\mu \nu }  
+ g_{\mu \nu } \;p }{2}\Big) + \frac{\kappa^{2}}{2} \,\Theta_{\mu \nu }    \Bigg].
\end{eqnarray}
These previous equations lead to the following expression
\begin{eqnarray}
&& \nabla^\mu \Theta_{\mu \nu } = \frac{-2}{(f_{\mathcal{T}}+ \kappa^2) } 
\Bigg\{ \nabla^\mu  \Big[\frac{T\,g_{\mu\nu}\,f_T}{2} + S^{\rho}_{\;\;\;\mu\nu}\; f_{T\mathcal{T}}\;
\partial_{\rho} \mathcal{T}  -
S^{\rho}_{\;\;\;\mu\nu}\;f_{TT}\; \partial_{\rho} T  - \frac{1}{4} g_{\mu \nu } \Big(T+ f \Big)\Big]  + \cr
&& \Big(\frac{\Theta_{\mu \nu }  
+ g_{\mu \nu } \;p }{2}\Big)\,
\nabla^\mu f_{\mathcal{T}} + \frac{f_{\mathcal{T}}}{2}\,\nabla^\mu ( g_{\mu \nu } \;p )- G_{\mu\nu} \nabla^\mu (1+ f_T)\Bigg\}.
\end{eqnarray}
  The $f(T,\mathcal{T})= 0+ f(\mathcal{T})$ gravity field equations  namely  \Big($f(T)=0$ or $f_{TT} =f_{T\mathcal{T}} =f_{T}= 0$ \Big) become 
\begin{eqnarray}
 \nabla_\mu \Theta^\mu_\nu = \frac{1}{2(f_{\mathcal{T}}+ \kappa^2) } \Bigg\{ 
 \frac{1}{4} \delta^\mu_\nu \nabla_\mu  f(\mathcal{T})   -
 \Big(\frac{\Theta^\mu_\nu  
+ \delta^\mu_\nu  \;p }{2}\Big)\,
\nabla_\mu f_{\mathcal{T}} - \frac{f_{\mathcal{T}}}{2}\,\delta^\mu_\nu  \nabla_\mu  \,p \Bigg\},
\end{eqnarray}
where we have used the barotropic equation of state $p = \omega\rho$. By fixing $\nu= 0$, one gets:
\begin{eqnarray}\label{afrik}
 \dot{\rho}+3H\rho\left(1+\omega\right) = \frac{-\dot{\rho}}{2(f_{\mathcal{T}}+ \kappa^2) }
 \Bigg\{\frac{1}{4}\left(1-3\omega\right)\;f_{\mathcal{T}}  + 
 \Big(\frac{1 +  \;\omega }{2}\Big)\rho\, \left(1-3\omega\right) f_{\mathcal{T}\mathcal{T}} + \frac{f_{\mathcal{T}}}{2}\,\omega \Bigg\}
\end{eqnarray}
To ensure cancellation of the divergence of the  energy-momentum tensor, we vanish the second member of the equation  Eq.(\ref{afrik}), and obtain
the following differential equation
\begin{eqnarray}
\frac{1}{2} f(\mathcal{T}) + f_{\mathcal{T}}\; \mathcal{T} \;\frac{(1-\omega)}{(1+\omega)}=0, \,\label{s1'},
\end{eqnarray}
 whose general solution reads
\begin{eqnarray}
f(\mathcal{T})= Q\; \mathcal{T}^{\frac{(1+3\omega)}{2(1+\omega)}} , \label{s1}
\end{eqnarray}
\begin{eqnarray}
f(T, \mathcal{T})= T -2\Lambda + Q\; \mathcal{T}^{\frac{(1+3\omega)}{2(1+\omega)}} , \label{s1}
\end{eqnarray}
We report here that $ Q$  is the integration constant.
At the moment,  we are pointing out the exact expression of the constant  $Q$ by using the wonderful conditions mentioned in
\cite{st28} 
which stipulates that the algebraic function $  f( \mathcal{T})=\mathcal{T}^{\mathcal{N}} $ must satisfy the following initial conditions
\begin{eqnarray}
\left(f\right)_{t=t_i}=T_i,\quad \left(\frac{df}{dt}\right)_{t=t_i}=\left(\frac{dT}{dt}\right)_{t=t_i},\label{arsene1}
\end{eqnarray} 
with $t_i$ the early time and $T_i$ the initial valor of the scalar torsion associated.
By making use of this initial condition  (\ref{arsene1}) and (\ref{s1}), one expresses the constant $Q$ as
\begin{eqnarray}
Q=2 \Lambda \; \mathcal{T}_0^{-\frac{(1+3\omega)}{2(1+\omega)}},\label{cvalue}
\end{eqnarray}
and the associated algebraic function is
\begin{eqnarray}
f(T, \mathcal{T})= T + 2\Lambda \Big[ \Big(\frac{\mathcal{T}}{\mathcal{T}_0}\Big)^{\frac{(1+3\omega)}{2(1+\omega)}}- 1 \Big] 
. \label{reconstruit}
\end{eqnarray}
 We emphasize here the constant $Q$ is positive because of the positivity of $\Lambda$ parameter. Moreover, if it vanishs ($ Q=0 $), we come back to  the TT equivalent of RG.
 \section{ Dynamic study of the systems}\label{sec4}
 We are working in this section whit  the cosmological flat metric of  $FLRW$ described by
\begin{eqnarray}
 ds^{2}= dt^{2} - a^{2}(t)\left(dx^2+dy^2+dz^2\right),  \label{metricflat}
 \end{eqnarray}
from which we obtein the diagonal matrix for the tetrads as
\begin{eqnarray}
 \{e^{a}_{\;\; \mu}\}= diag[1,a,a,a]. \label{eq11}
\end{eqnarray}
 The determinant of the matrix (\ref{eq11}) is $e = a^{3}$ and the non zero components of torsion and contorsion are given by
\begin{eqnarray}
 T^{1}_{\;\;\; 01}= T^{2}_{\;\;\; 02}=T^{3}_{\;\;\; 03}=\frac{\dot{a}}{a},\\
 K^{01}_{\;\;\;\;1}=K^{ 02}_{\;\;\;\;2}=K^{ 03}_{\;\;\;\;3}= \frac{\dot{a}}{a},  \label{eq12}
\end{eqnarray}
The calculus of components of $S^{\;\;\; \mu\nu}_{\alpha}$ also gives:
\begin{eqnarray}
S^{\;\;\; 11}_{0}=S^{\;\;\; 22}_{0}=S^{\;\;\; 33}_{0}=\frac{\dot{a}}{a}.  
\end{eqnarray}
Therefore, the scalar torsion is expressed as 
\begin{eqnarray}
 T= -6H^{2}, \label{m1}
\end{eqnarray}
where $H=\dot{a}/a$ denotes the Hubble parameter. We report also the expression of the trace of energy-momentum tensor related to matter,
$ \Theta =\mathcal{T}= (1 - 3 \omega) \rho$. We assume now that the ordinary component of Universe is a perfect fluid with the equation of state 
 $p = \omega \rho$  and  $c^2_s= \dot{p}/\dot{\rho}$  so that the energy-momentum is given by 
\begin{eqnarray}
 \Theta_{ \mu \nu} =
diag(1 , -\omega, -\omega,
-\omega ) \rho. 
\end{eqnarray}
To point out an application of this theory in Cosmology, we insert as needful the flat metric of FLRW  (\ref{metricflat}) in the field equations
\eqref{lagran1}; and obtain consequently the  Friedmann modified equations  below
\begin{eqnarray}
H^{2}&=&\frac{8\pi G}{3} \rho-\frac{1}{6}\left(f+12 H^{2} f_{T}\right)+f_{%
\mathcal{T}} \left(\frac{\rho+p}{3}\right),\cr
\dot{H}&=&-\frac{4\pi G\left( 1+f_{\mathcal{T}}/8\pi G\right) \left( \rho+p\right) }{1+f_{T}-12H^{2}f_{TT}+H\left( d\rho /dH\right)
\left( 1-3c_{s}^{2}\right)f_{T\mathcal{T}} }.
\end{eqnarray}
 where $\rho=\rho_m +\tilde{\rho} +\rho_r$, while $\rho_m$, $ \tilde{\rho}$ and
 $\rho_r$ represent the energy densities of matter, dark energy and
 the radiation respectively. We also suppose that these three components of the above defined fluid are in interactions. The  continuity equations  taking into account the different interactions  are written as
 \begin{eqnarray}
\dot{\rho_m}+3H(\rho_m+p_m)&=& E_1, \label{1}\cr
\dot{\tilde{\rho}}+3H(\tilde{\rho}+\tilde{p})&=& E_2, \label{2}\cr
\dot{\rho_r}+3H(\rho_r+p_r) &=& E_3, \label{1'}
\end{eqnarray}
 where  $E_i$, $i=1,2,3$  are  the term of interaction between the two fluids. Now, we can define the cosmological density parameters
\begin{eqnarray}
y= \frac{\kappa^2 \rho_m}{3H^{2}},\quad x= \frac{\kappa^2\tilde{\rho}}{3H^{2}}.\quad z= \frac{\kappa^2\rho_r}{3H^{2}}
\label{a6}
\end{eqnarray}
and
\begin{eqnarray}
\frac{\dot{H}}{H^2}
=-\frac{3/2 \left( 1+f_{\mathcal{T}}/8\pi G\right) \left(x+y+z+ x\,\tilde{\omega} +\omega_r\,z \right) }{1+f_{T}-12H^{2}f_{TT}+H\left( d\rho/dH\right)
\left( 1-3c_{s}^{2}\right)f_{T\mathcal{T}} }.
\end{eqnarray}
By using the  $e$-folding parameter, $ Z= \ln{a}$, $a$ being the scale factor, the continuity equations Eqs. (\ref{1}) and (\ref{2}) become
\begin{eqnarray}\label{mouss}
\frac{dx}{dZ}&=& 3x \Bigg( \frac{\left( 1+f_{\mathcal{T}}/8\pi G\right) \left(x+y+z+ x\,\tilde{\omega} +\omega_r\,z \right) }{1+f_{T}-12H^{2}f_{TT}+H\left( d\rho/dH\right)
\left( 1-3c_{s}^{2}\right)f_{T\mathcal{T}} }      \Bigg) -3x(1 + \tilde{\omega})  +       \frac{\kappa^2\,E_1}{3H^3} \label{3}\cr
\frac{dy}{dZ}&=& 3y \Bigg( \frac{\left( 1+f_{\mathcal{T}}/8\pi G\right) \left(x+y+z+ x\,\tilde{\omega} +\omega_r\,z \right) }{1+f_{T}-12H^{2}f_{TT}+H\left( d\rho/dH\right)
\left( 1-3c_{s}^{2}\right)f_{T\mathcal{T}} }      \Bigg) -3y   +     \frac{\kappa^2\,E_2}{3H^3} \label{4}\cr
 \frac{dz}{dZ}& =&   3z \Bigg( \frac{\left( 1+f_{\mathcal{T}}/8\pi G\right) \left(x+y+z+ x\,\tilde{\omega} +\omega_r\,z \right) }{1+f_{T}-12H^{2}f_{TT}+H\left( d\rho/dH\right)
\left( 1-3c_{s}^{2}\right)f_{T\mathcal{T}} }      \Bigg) -3z(1 + \omega_r)  +       \frac{\kappa^2\,E_3}{3H^3}, 
\end{eqnarray}
  Where we have used unit of $ \kappa^{2}=1$ and then $Z\equiv N\equiv\ln a$ is used as e-folding parameter. The interacting  parameters 
  $E_i$, $i=1,2,3$ are generally functions of the energy densities  and the Hubble parameter i.e $E_i= E_i(H,\rho_i)$. We start the analysis of the system
  of equations  in (\ref{mouss}) by vanishing the first member of each of these equations  in order to extract critical points. Therefore, one perturbs
   these equations  in first order around the critical points and deduce the stability of the system. In our calculation procedure, we force the following
   parameters $\omega_m=0$, $\omega_r=\frac{1}{3}$ and $\tilde{\omega}$ to be non zero but negative. We are interested to the stable critical points i.e
    the points for which the  eigenvalues of Jacobian matrix associated to the system of equations  are negative. Such of points are useful because they
    represent the attractive solutions of  dynamic system. 
 
\section{Analysis of stability in phase space}

In this section, we will erect four models by choosing different forms of coupling parameters  $E_i$  and we will analysis the stability of the 
corresponding  dynamic system around the critical points and  plot the evolutionary phase diagram associated. To reach this target, we must search for 
the critical points of (\ref{mouss}) and make the system linear around the above points.

 \subsection{Interacting model - I}

 We consider the models with  the following interaction terms

\begin{equation}\label{8a}
E_1=-6bH\tilde{\rho}, \ \ E_2=E_3=3bH\tilde{\rho},
\end{equation}

 where $b$ is a coupling parameter assumed to be  positive real in the input parameters. Then, the equation Eq. (\ref{8a}) shows that matter and 
 the radiation have energy densities  which increase with the time whereas the energy density of dark energy is going to disappears completely. So, 
 the dar energy declines for matter and radiation.

Using (\ref{reconstruit}) and   (\ref{8a}), the system (\ref{mouss}) takes the form

\begin{eqnarray}
\frac{dx}{dN}&=&-3x(1+\tilde{\omega})+3x\Big(x+y+z+\tilde{\omega} x+\omega_r z\Big)-6bx,\nonumber\\
\frac{dy}{dN}&=&-3y+3y\Big(x+y+z+\tilde{\omega} x+\omega_r z\Big)+3bx,\label{33}\\
\frac{dz}{dN}&=&-3z(1+\omega_r)+3z\Big(x+y+z+\tilde{\omega} x+\omega_r
z\Big)+3bx.\nonumber
\end{eqnarray}
 \begin{figure}[h]
   \centering
   \begin{tabular}{rl}
   \includegraphics[width=7cm, height=5cm]{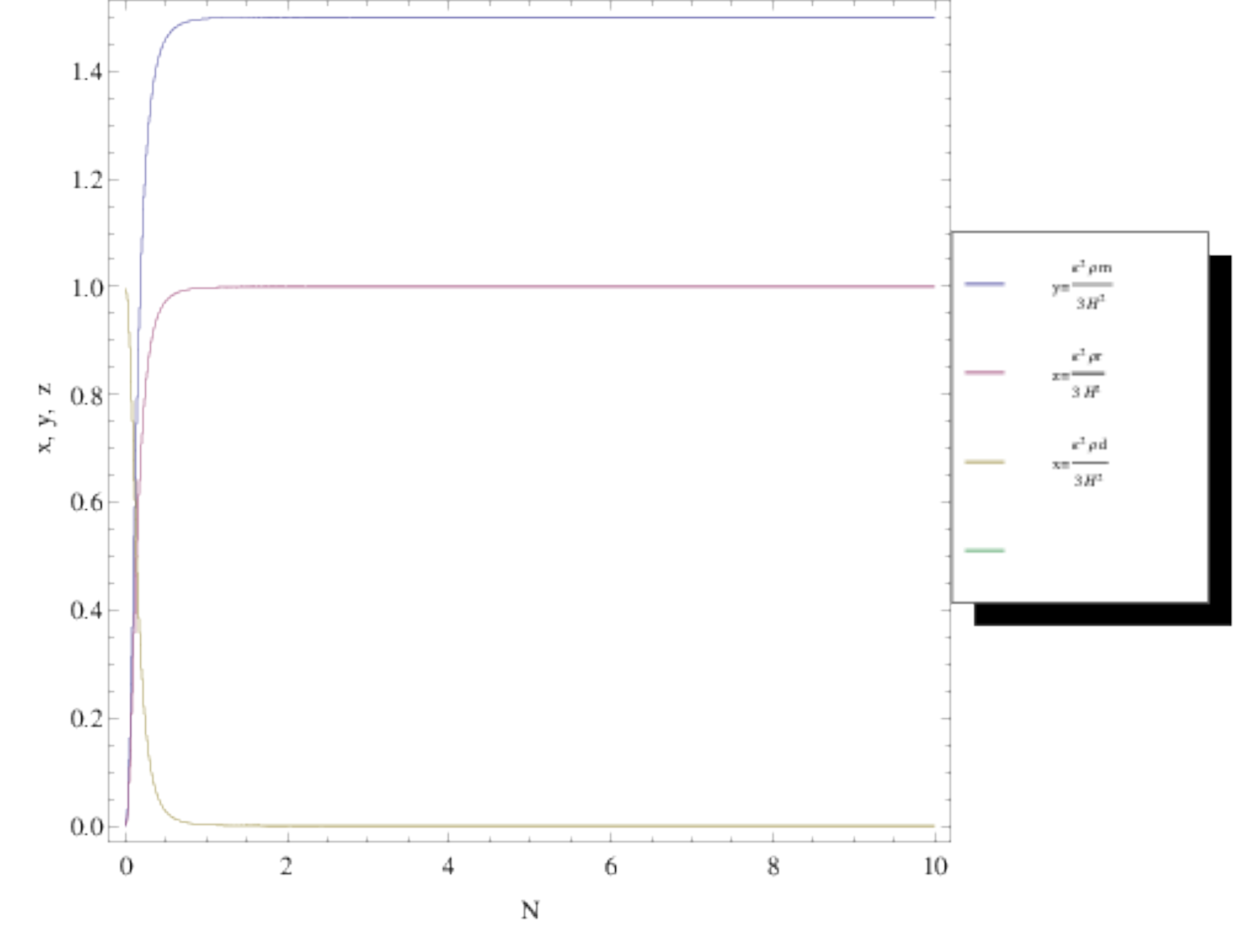}
   \end{tabular}
   \caption{
  Model I: variation of $x,y,z$ as a function of the $N=\ln(a)$. The initial
   conditions chosen are $x(0)=0.7,y(0)=0.3,z(0)=0.01$,
   $\tilde{\omega}=-1.2,\,\,\omega_r=\frac{1}{3}$ and $b=0.5$. 
   }
  \label{fig1}
   \end{figure}

 The critical points are found for this model by vanishing the first member of (\ref{33}) and there are the following four points recorded in the 
 below board. 

 \begin{widetext}
   \begin{table}[ht]
   \centering
   \begin{tabular}{|c| c |c| c |c|}
   \hline\hline
     Point & $(x_c,y_c,z_c)$ & $\lambda_1$ & $\lambda_2$ & $\lambda_2$ \\ [0.5ex]
      \hline
      $ P_{11}$ & $(0,0,0)$ &$3(b-1)$ & $ -3(b-1)$ & $-3(1+2b+\tilde{\omega})$ \\\hline
      $ P_{12}$ & $(0,(1-b),0)$ &$-1 $& $ 3(1-b) $ & $ -3(\tilde{\omega}+3b) $\\ \hline
      $ P_{13}$ &$(0,0,\frac{3}{4}(\frac{4}{3}-b))$  & $1$& $ 4-3b $ & $ 1-3\tilde{\omega}-9b  $\\ \hline
       $P_{14}$ &$(\frac{(1+\tilde{\omega}+2b)}{1+\tilde{\omega}},0,0)$&$3(1+\tilde{\omega}+2b)$& $ 3(\tilde{\omega}+3b)$ &
       $ -1+3(\tilde{\omega}+3b)$\\ \hline
       \end{tabular}
         \caption{Critical points and The eigenvalues for the first model}
        \label{table1}
        \end{table}
 \end{widetext}

 The point  $P_{11}$ is stable when one the following conditions is satisfied
\begin{eqnarray}
\tilde{\omega}<-3,\ \ b<1/18.\\
\tilde{\omega}\geq-3, \ \ \tilde{\omega}<-10/9,\ \ b<1/18.\\
\tilde{\omega}\geq-\frac{10}{9},\ \ \tilde{\omega}<0,\ \  b<-\frac{1}{2}(1+\tilde{\omega}).
\end{eqnarray}
 
 $ P_{12}$ is an unstable critical point because even if $\lambda_1<0$ we have  $\lambda_2<0$ si $b>1$.  $ P_{13}$ is stable if 
 $b>1,\tilde{\omega}>-3b$. In parallel $P_{14}$ is not stable because  $\lambda_1>0$.\par

 \begin{figure}[h]
   \centering
   \begin{tabular}{rl}
   \includegraphics[width=8cm, height=6cm]{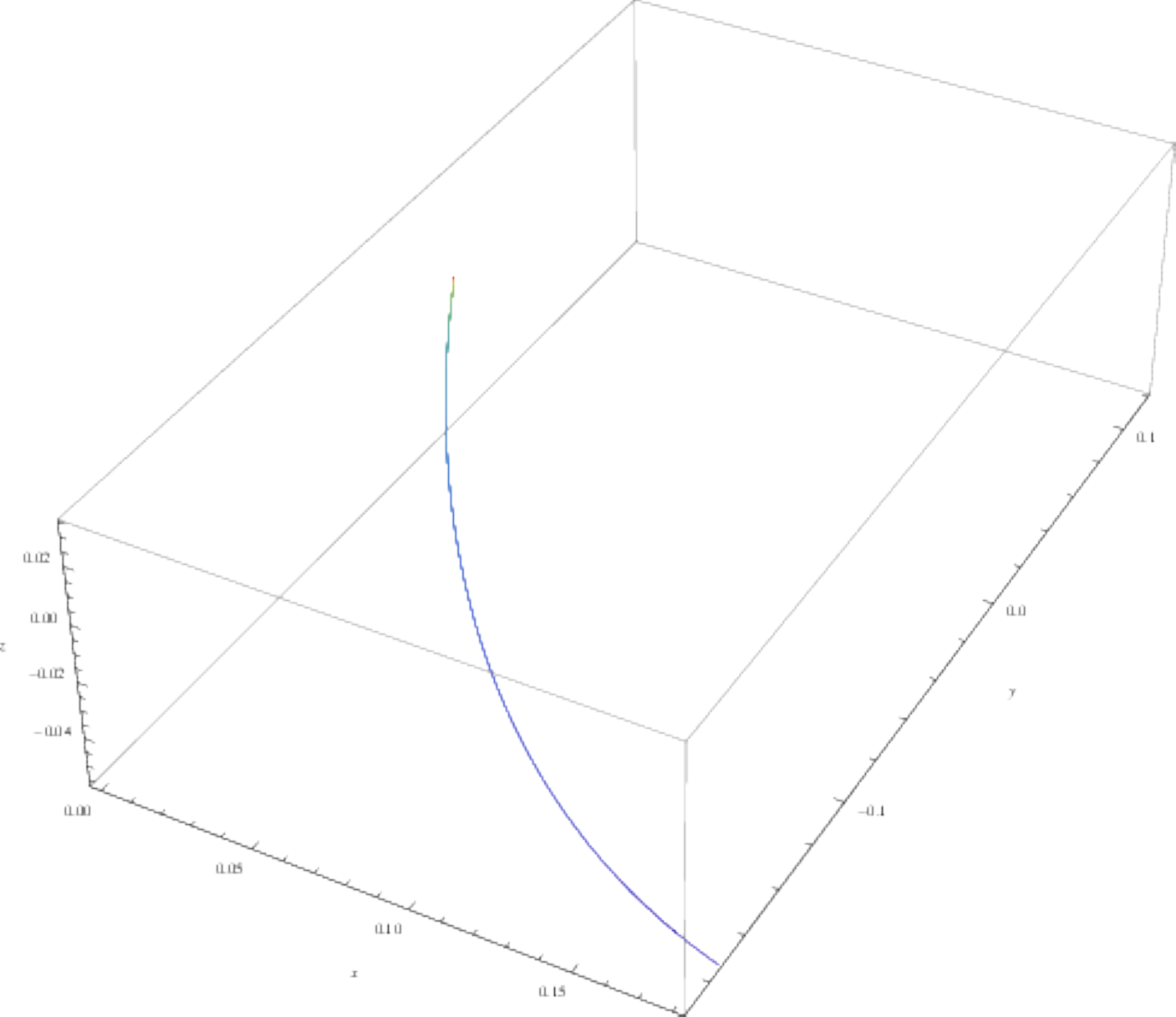}&
    \includegraphics[width=8cm, height=6cm]{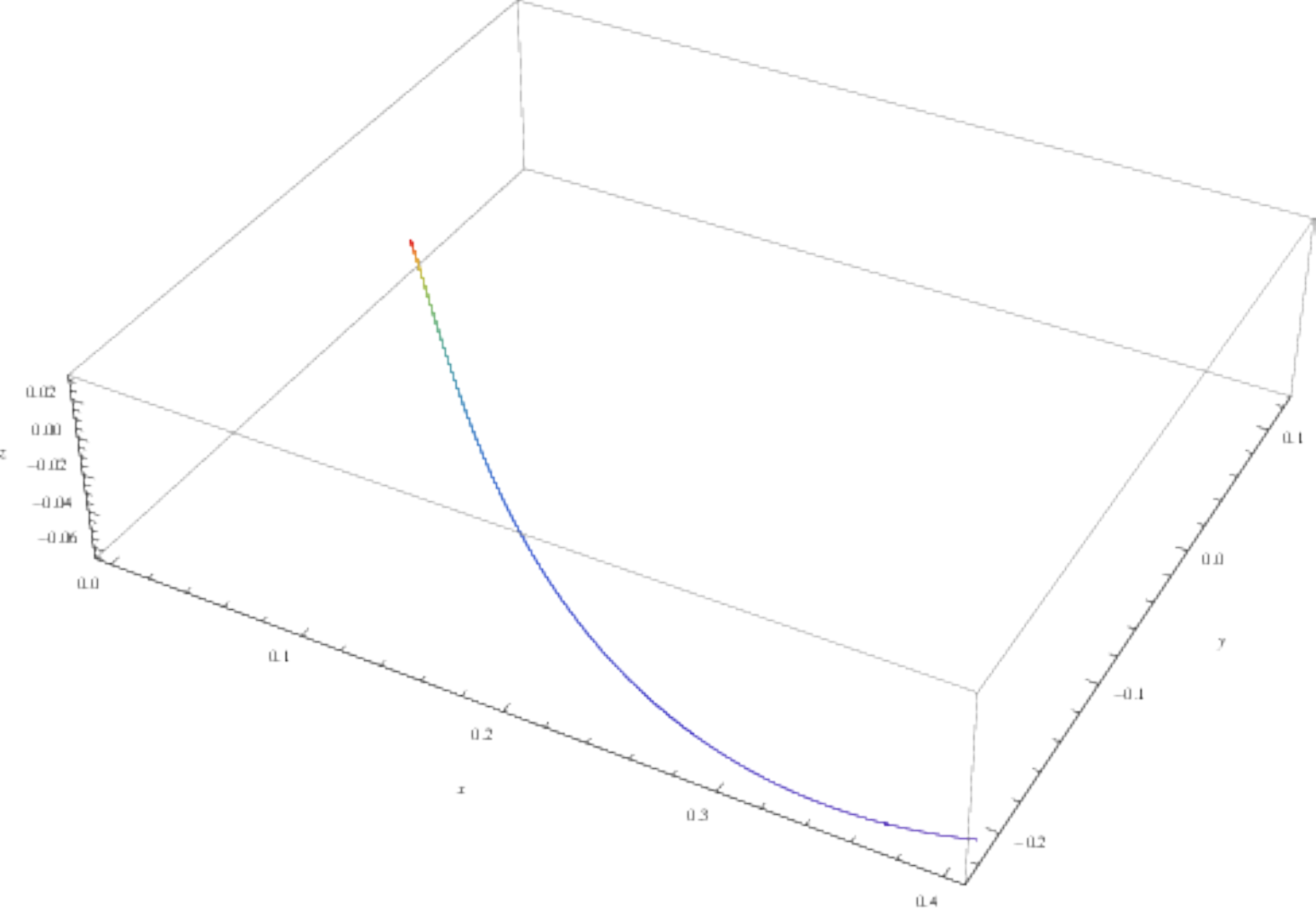}\\
     \includegraphics[width=8cm, height=6cm]{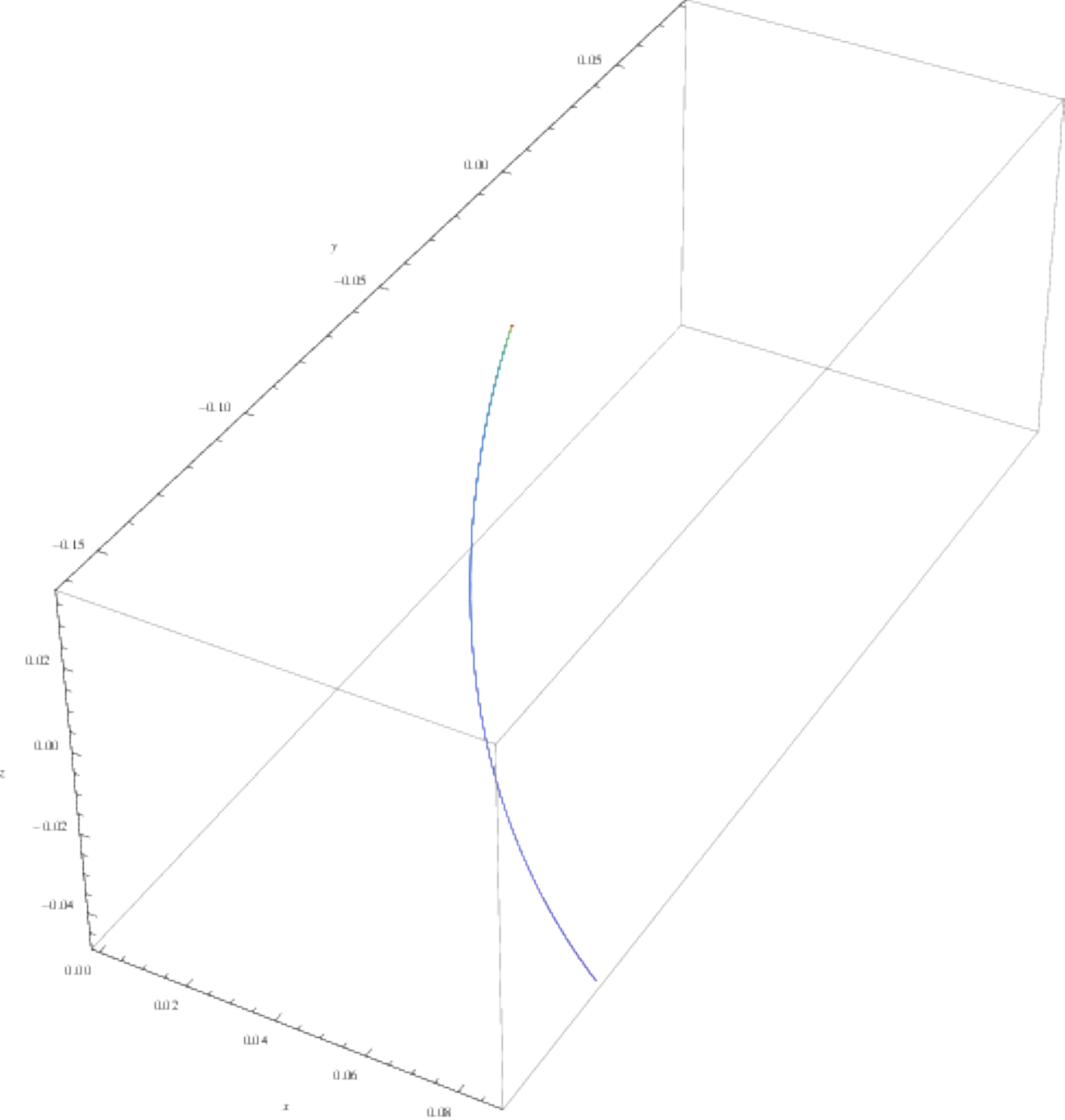}
   \end{tabular}
   \caption{ Model I:  Phase space for $\tilde{\omega}=(-1,-1.2,-1.5),\, b=0.5,\,\omega_r=\frac{1}{3}$}
  \label{fig2}
   \end{figure}

It follows that the matter density dominates for the model {\it I} whose parameters stay for the conditions $-1<w_d<-\frac{1}{3},w_u>0,b=0.5$, $w_{tot}>0$.
 This conclusion is confirmed by the fig.\ref{fig1} where the matter density dominates whereas the radiation density is above the dark energy 
 density. The fig \ref{fig2} shows the phase diagram of the interaction between  dark energy and the both matter and radiation. According to
 the model  {\it I}, the dark energy density behaves like quintessence while matter and radiation densities fall with expansion.

\subsection{Interacting model - II}
We study another model with the choice of the interaction terms under the following form
\begin{equation}\label{9a}
E_1=-3bH\tilde{\rho},\ \ E_2=3bH(\tilde{\rho}-\rho_m),\ \
E_3=3bH\rho_m.
\end{equation}
 
 This model shows indeed the situation in which the dark energy looses his density in favor of  the matter whereas the radiation density increases
  because of its interaction with the matter.  

\begin{eqnarray}\label{9}
\frac{dx}{dN}&=&3x\Big(x+y+z+\tilde{\omega} x+\omega_r z \Big)-3bx-3x(1+\tilde{\omega}),\cr
\frac{dy}{dN}&=&3y\Big(x+y+z+\tilde{\omega} x+\omega_r z  \Big) + 3b(x-y)-3y,\cr
\frac{dz}{dN}&=&3z\Big(x+y+z+\tilde{\omega} x+\omega_r z\Big) + 3by-3z(1+\omega_r),
\end{eqnarray}

 \begin{figure}[h]
   \centering
   \begin{tabular}{rl}
   \includegraphics[width=7cm, height=5cm]{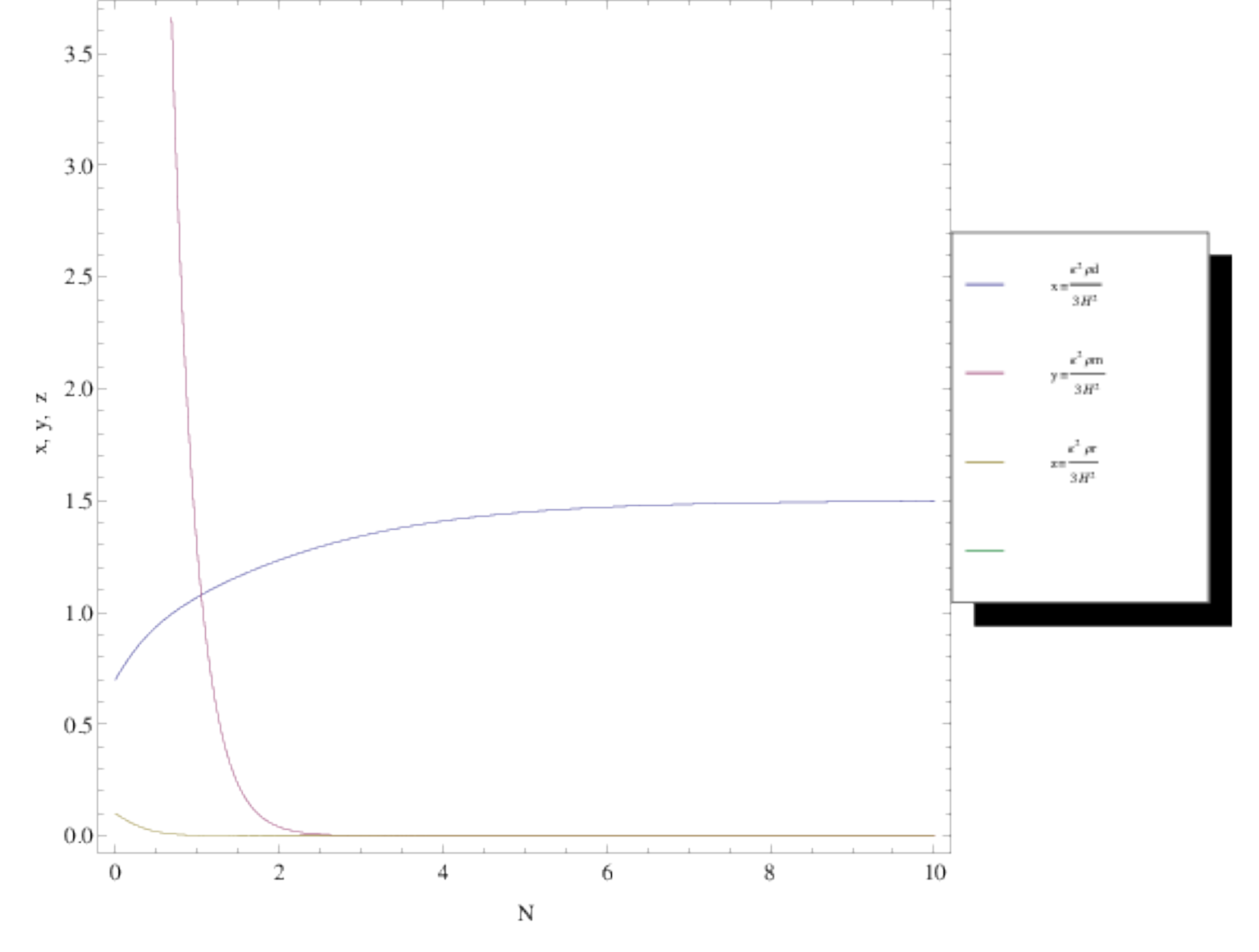}
   \end{tabular}
   \caption{
  Model I: variation of $x,y,z$ as a function of the $N=\ln(a)$. The initial
   conditions chosen are $x(0)=0.7,y(0)=0.3,z(0)=0.01$,
   $\tilde{\omega}=-1.2,\,\,\omega_r=\frac{1}{3}$ and $b=0.5$. 
   }
  \label{fig3}
   \end{figure}

We have free critical points

 \begin{tabular}{|c|l|p{2cm}|m{2.5cm}|b{6.5cm}|}
 \hline 
 Points & $\lambda_1$&$\lambda_2$ &$\lambda_3$ &  $(x_c,y_c,z_c)$
 \\\hline
 $ P_{21}$ &$3(b-1)$ & $ -3(b-1)$& $-3(1+2b+\tilde{\omega})$ &  $(0,\quad 0,\quad 0)$
 \\\hline
 $ P_{22}$ & $-1$ &$3(1-b)$&$ -3(\tilde{\omega}+3b) $ & $(0,\quad0,\quad 1)$
 \\\hline
  $ P_{23}$   & $1$& $ 4-3b $ & $ 1-3\tilde{\omega}-9b  $ & $\Big(0,\quad (1-3b),\quad 3b \Big)$
 \\\hline
 $ P_{24}$ &$3(1+\tilde{\omega}+2b)$& $ 3(\tilde{\omega}+3b)$ &
       $ -1+3(\tilde{\omega}+3b)$&
 $    \Bigg( -\,{\frac {3 \left( 2+\tilde{\omega} \right)  \left( b-\tilde{\omega}-1 \right)
  \left( b-\tilde{\omega}-7/3 \right)   }{3\,{\tilde{\omega}}^
 {3}+ \left( 16-3\,b \right) {\tilde{\omega}}^{2}+ \left( -6\,b+27 \right)
 w_{ {d}}+14+{b}^{2}+b}}, \newline
 -\,{\frac {b 3 \left( b-\tilde{\omega}-7/3 \right)
  \left( b-\tilde{\omega}-1 \right) }{3\,{\tilde{\omega}}^{3}+ \left( 16-3\,b
  \right) {\tilde{\omega}}^{2}+ \left( -6\,b+27 \right) \tilde{\omega}+14+{b}^{2}+b}}, \newline
  \,{\frac { 3  \left( b-\tilde{\omega}-1 \right) {b}^{2
 }}{14+16\,{\tilde{\omega}}^{2}-6\,\tilde{\omega}b+{b}^{2}-3\,{\tilde{\omega}}^{2}b+3\,{w_{{
 d}}}^{3}+27\,\tilde{\omega}+b}}
  \Bigg)$
 \\\hline
 \end{tabular}

  \begin{figure}[h]
  \centering
  \begin{tabular}{rl}
  \includegraphics[width=8cm, height=6cm]{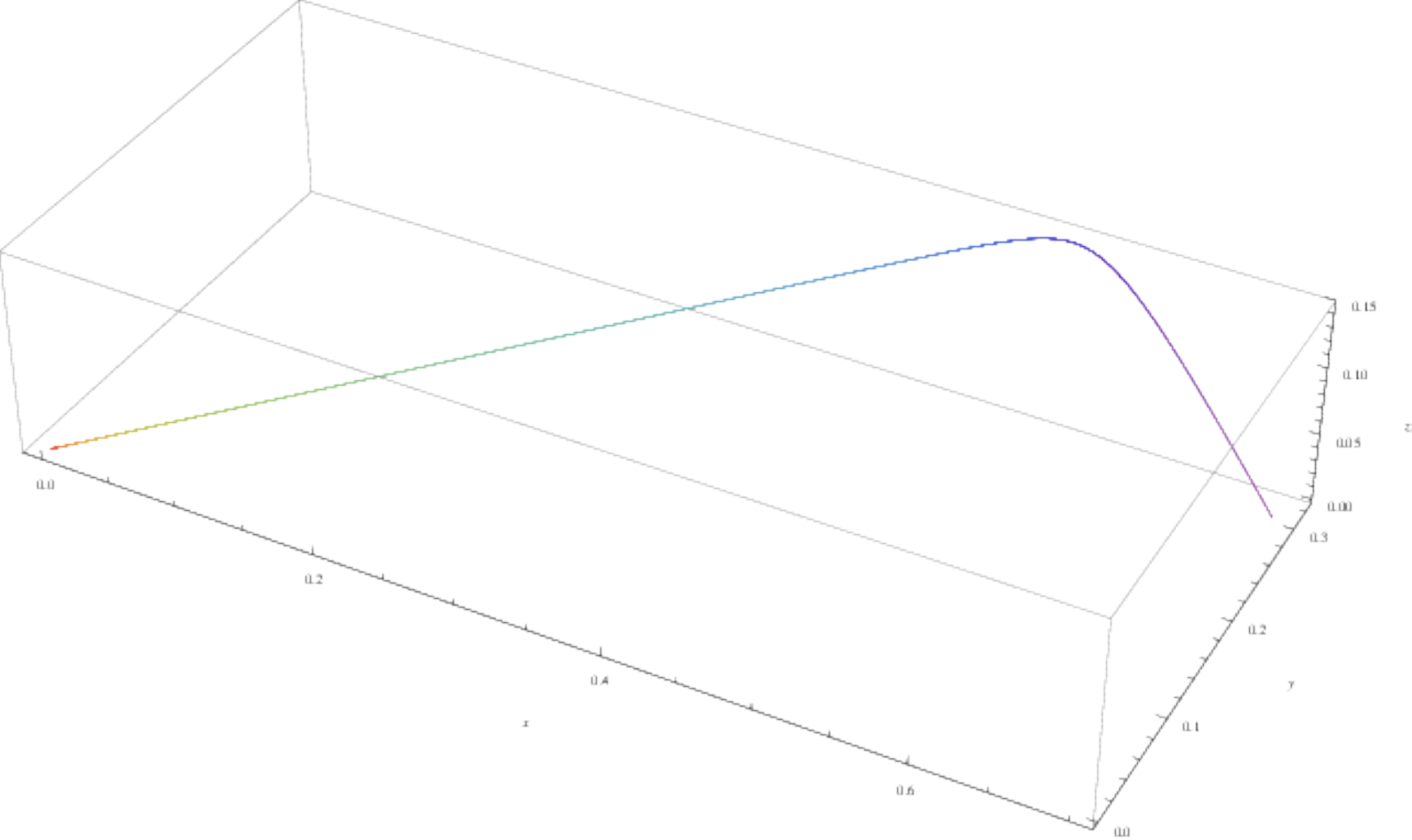}&
   \includegraphics[width=8cm, height=6cm]{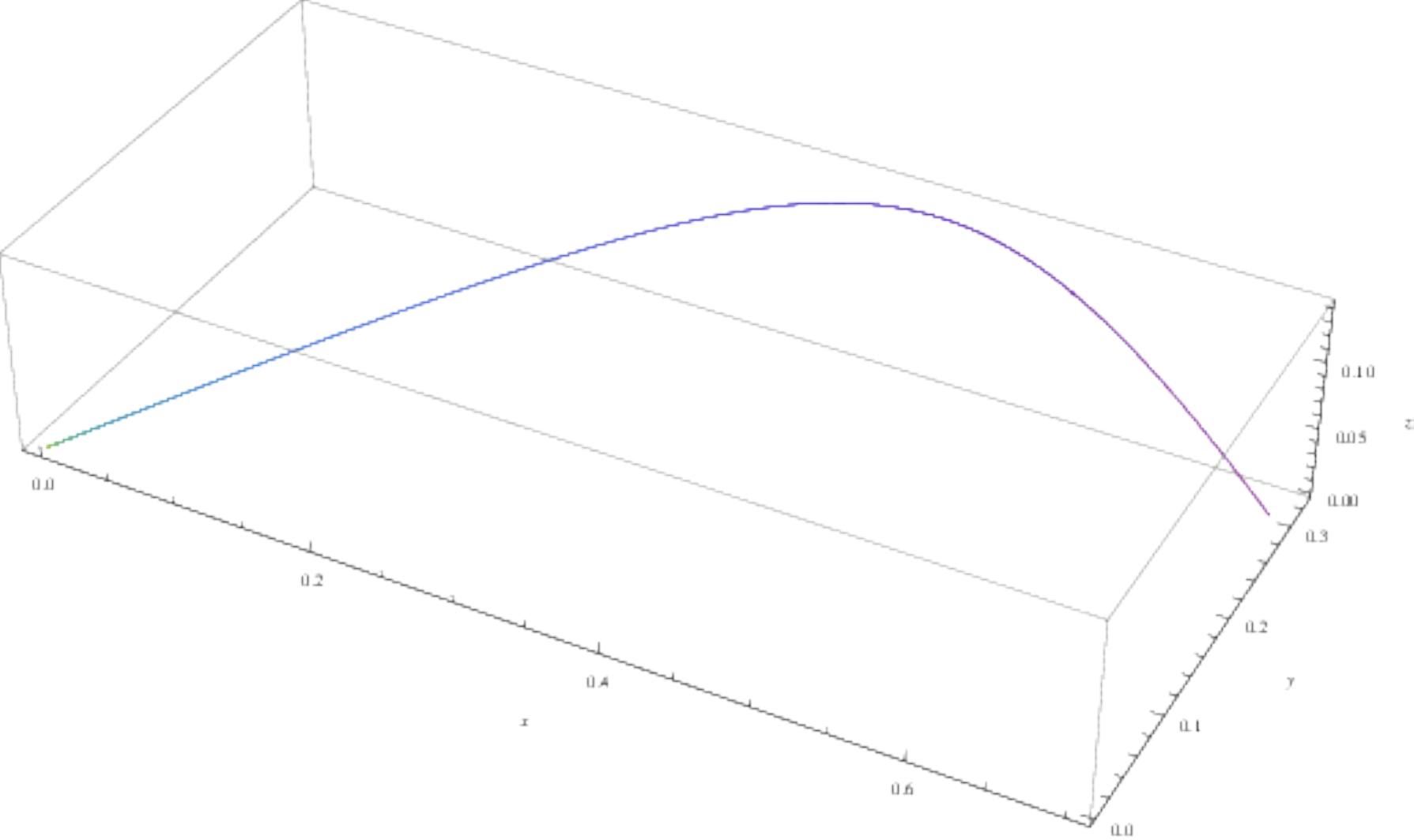}\\
    \includegraphics[width=8cm, height=6cm]{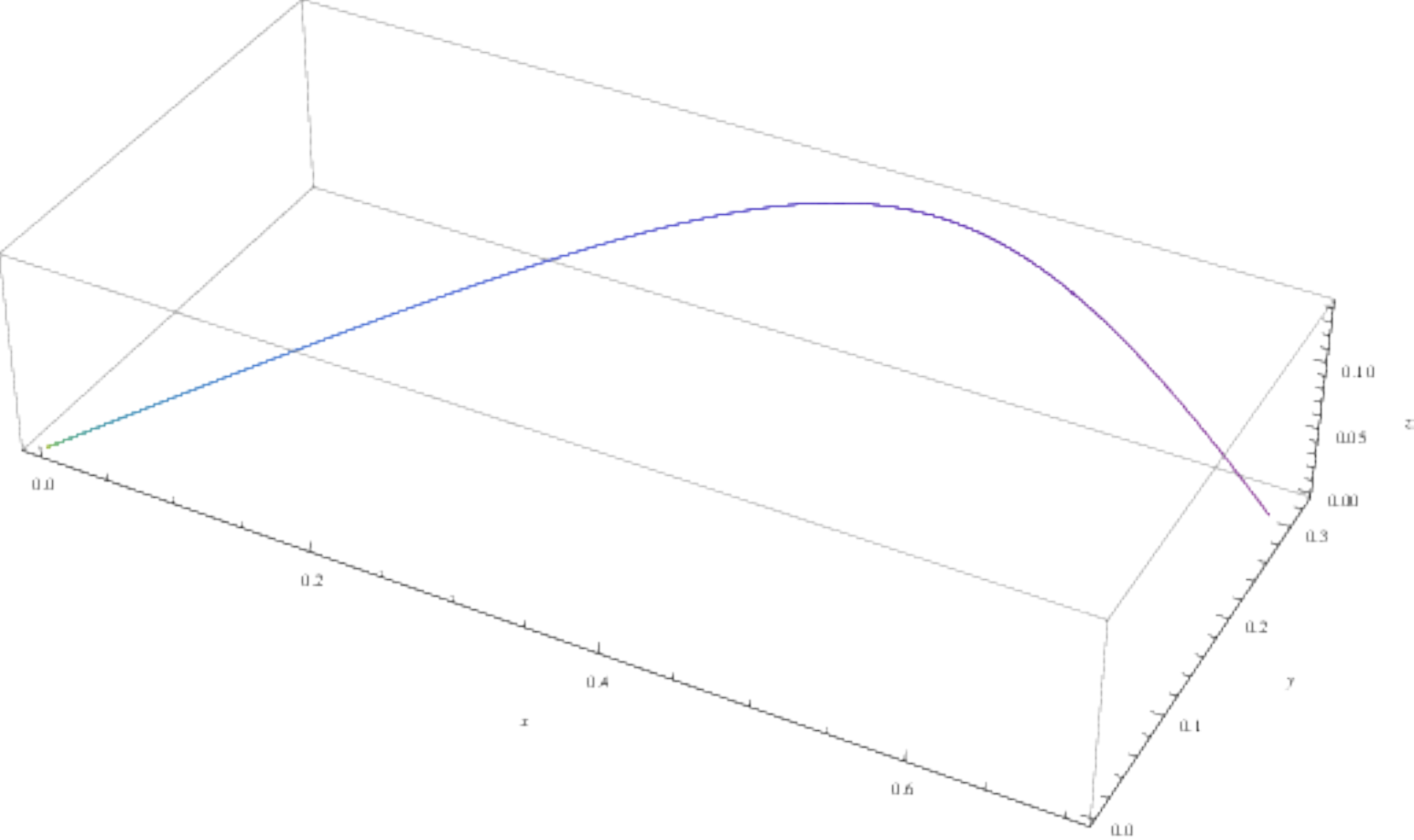}
  \end{tabular}
  \caption{ Model II:  Phase space for $\tilde{\omega}=(-1,-1.2,-1.5),\, b=0.5,\,\omega_r=\frac{1}{3}$}
 \label{fig4}
  \end{figure}
 
  \newpage

$ P_{21}$,  $ P_{24}$  are conditionally stable if  $b>1+\tilde{\omega}$ (for $ P_{21}$) and $\tilde{\omega}>-2$  and then 

$b<1+\tilde{\omega}$   ( for $ P_{24}$ ). But   $ P_{22}$ and   $ P_{23}$ are  unstable  because  $\lambda_1>0$0. 

  The figures  \ref{fig3} and  \ref{fig4} show the dynamic of the model {\it II}. We notice for this model that there is a great domination of dark
  energy while the energy densities of the radiation and the matter  have declined considerably. This situation is well compatible with the recent 
  observational data which show that the dark energy is the very important  responsible of the expansion of Universe.
  We also point out from these figures that if 
 $N\sim2$, the radiation declines and goes towards zero.  The figure \ref{fig4} is related to the phase diagram of radiation and dark energy interaction.
  For the model in study, the behavior of the dark energy is similar to quintessence while the matter and radiation tumble during the expansion. 
 
 \subsection{Interacting Model - III}
  Let us take the following interaction terms \cite{jamil8}

\begin{equation}\label{12}
E_1=-6b\kappa^2 H^{-1}\tilde{\rho}\rho_r, \ \
E_2=E_3=3b\kappa^2 H^{-1}\tilde{\rho}\rho_r.
\end{equation}
The system in (\ref{mouss}) becomes
\begin{eqnarray}
\frac{dx}{dN}&=&3x\Big(x+y+z+\tilde{\omega}x+\omega_r z \Big)
-3x-3\tilde{\omega}x-18bxz,\nonumber\\
\frac{dy}{dN}&=&3y\Big(x+y+z+\tilde{\omega}x+w_ rz\Big)-3y+9bxz,\\
\frac{dz}{dN}&=&3z\Big(x+y+z+\tilde{\omega}x+\omega_r
z\Big)-3z-3\omega_rz+9bxz.\nonumber
\end{eqnarray}
.
 
 \begin{figure}[h]
   \centering
   \begin{tabular}{rl}
    \includegraphics[width=8cm, height=6cm]{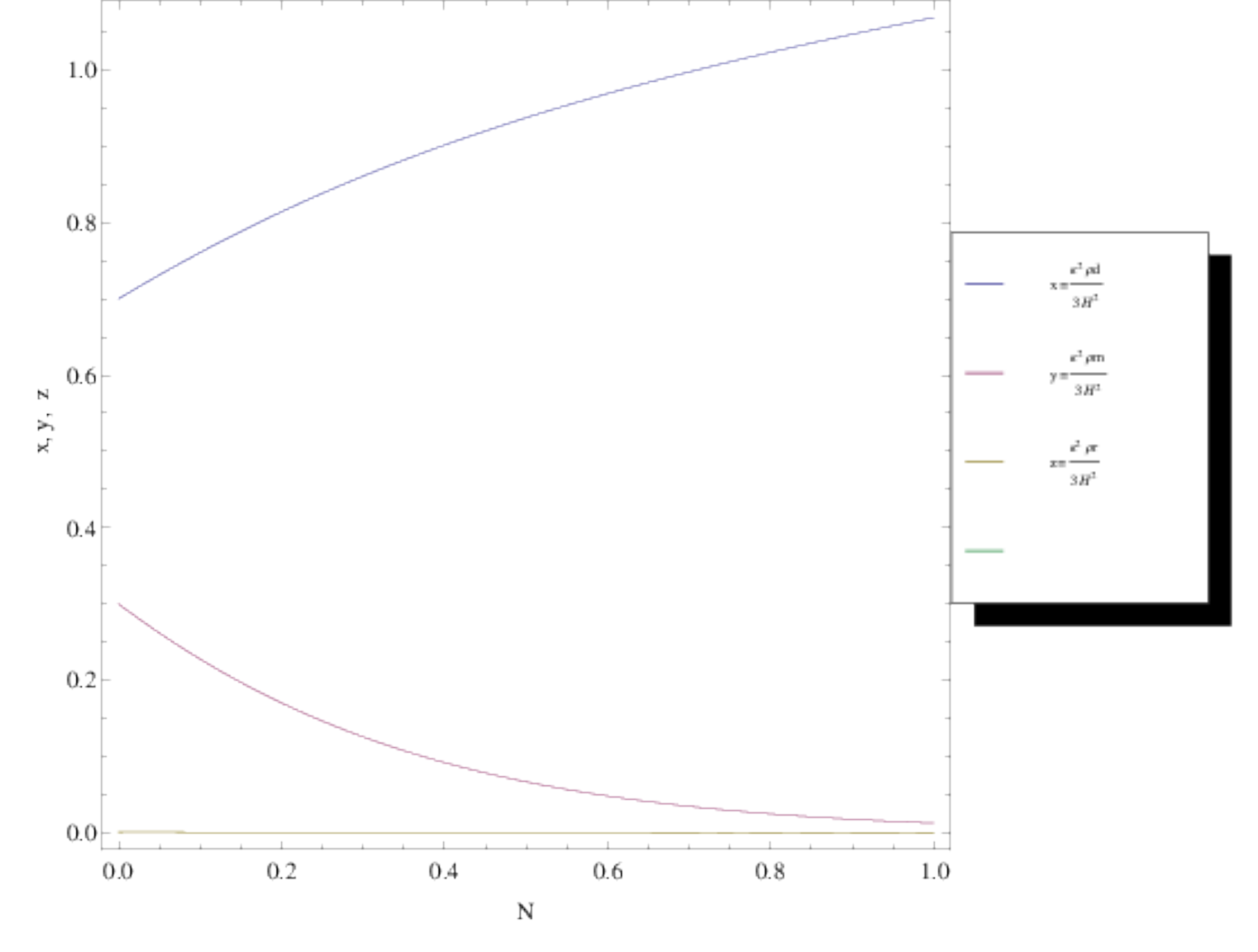}
   \end{tabular}
   \caption{ Model III:   variation of $x,y,z$ as a function of the $N=\ln(a)$. The initial
   conditions chosen are $x(0)=0.7,y(0)=0.3,z(0)=0.01$,
   $\tilde{\omega}=-1.2,\,\,\omega_r=\frac{1}{3}$ and $b=0.5$.}
  \label{fig5}
   \end{figure}
 
 \begin{tabular}{|c|l|p{3cm}|m{3cm}|b{6cm}|}
 \hline 
 Points & $\lambda_1$&$\lambda_2$ &$\lambda_3$ &  $(x_c,y_c,z_c)$
 \\\hline
 $ P_{31}$ &$-3$ & $ -4$& $-3(1+\tilde{\omega})$ &  $\ (0,\quad 0,\quad0)$
 \\\hline
 $ P_{32}$ & $3$ &$-1$&$ -3\tilde{\omega} $ & $\ (0,\quad 1,\quad0)$
 \\\hline
  $ P_{33}$   & $ 3\tilde{\omega} $& $ 3(1+\tilde{\omega}) $ & $ +9\,b-1+3\,\tilde{\omega} $ & $\ (0,\quad 0,\quad 1) $
 \\\hline
  $ P_{34}$   & $4$& $ 1-9b $ & $ -9b+1-3\tilde{\omega} $ & $ \ (1,\quad0,\quad0)$
 \\\hline
  $ P_{35}$   & -& - & - &
 \\\hline
 $ P_{35}$ &-& -&
       -&  $\Bigg( 
 {\frac { \left( \tilde{\omega}+6\,b-\frac{1}{3} \right) }{
 b \left(+6\,b-\frac{1}{3}+2\,\tilde{\omega} \right)
 }},\newline
 {\frac {\,-3\,b+18\,{b}^{2}+{\tilde{\omega}}^{2}-2\,\tilde{\omega} \frac{1}{3}+9\,\tilde{\omega}b+\frac{1}{9}}{3b \left(6\,b-\frac{1}{3}+ 2\,\tilde{\omega} \right) }},\newline
 -\,{\frac
 {\tilde{\omega} \left(+3\,b-\frac{1}{3}+\tilde{\omega} \right) } {3b
 \left(6\,b-\frac{1}{3}+2\,\tilde{\omega} \right) }}
 \Bigg)$
 \\\hline
 \end{tabular}
   
 \begin{figure}[h]
   \centering
   \begin{tabular}{rl}
   \includegraphics[width=8cm, height=6cm]{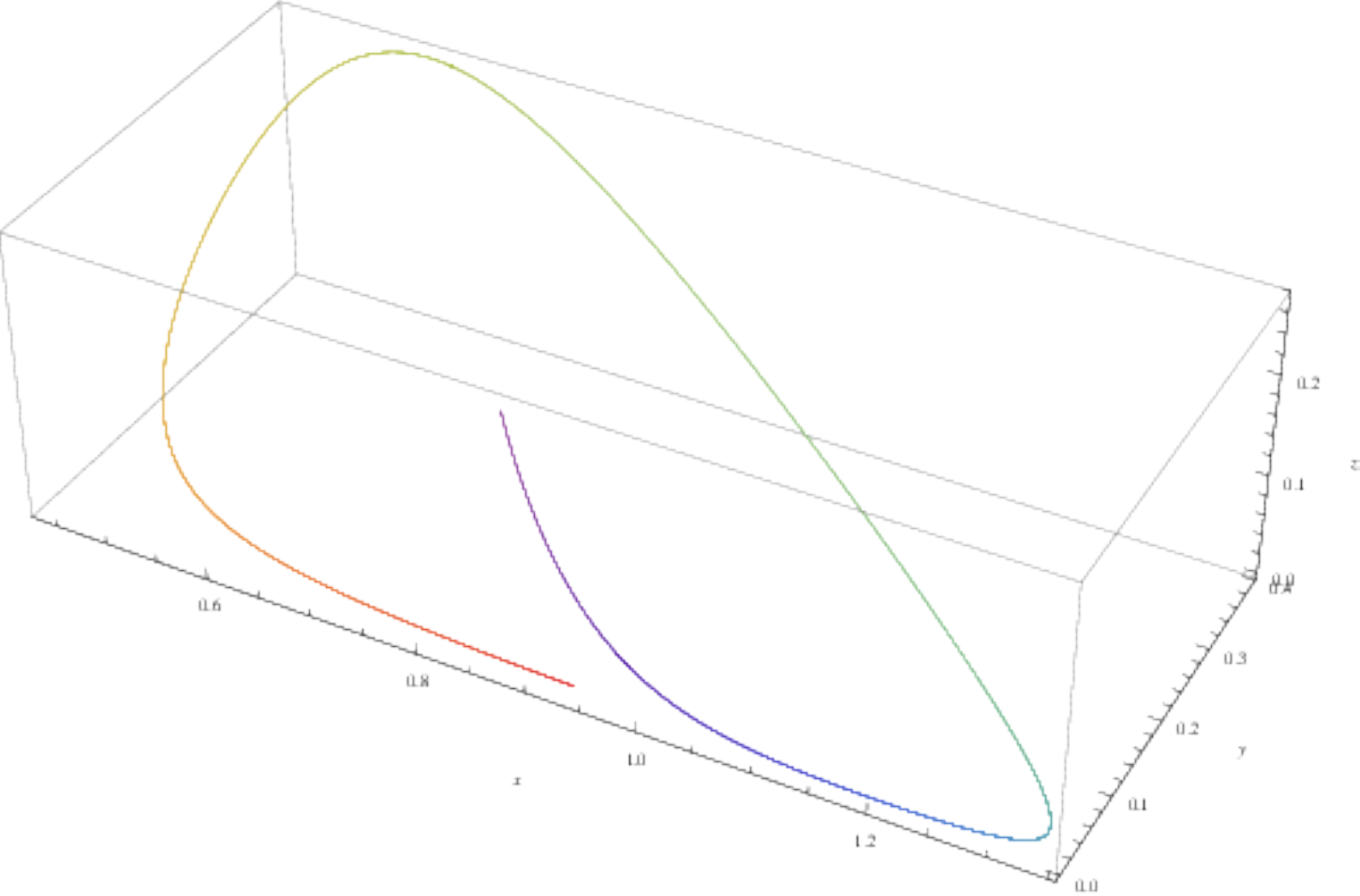}&
    \includegraphics[width=8cm, height=6cm]{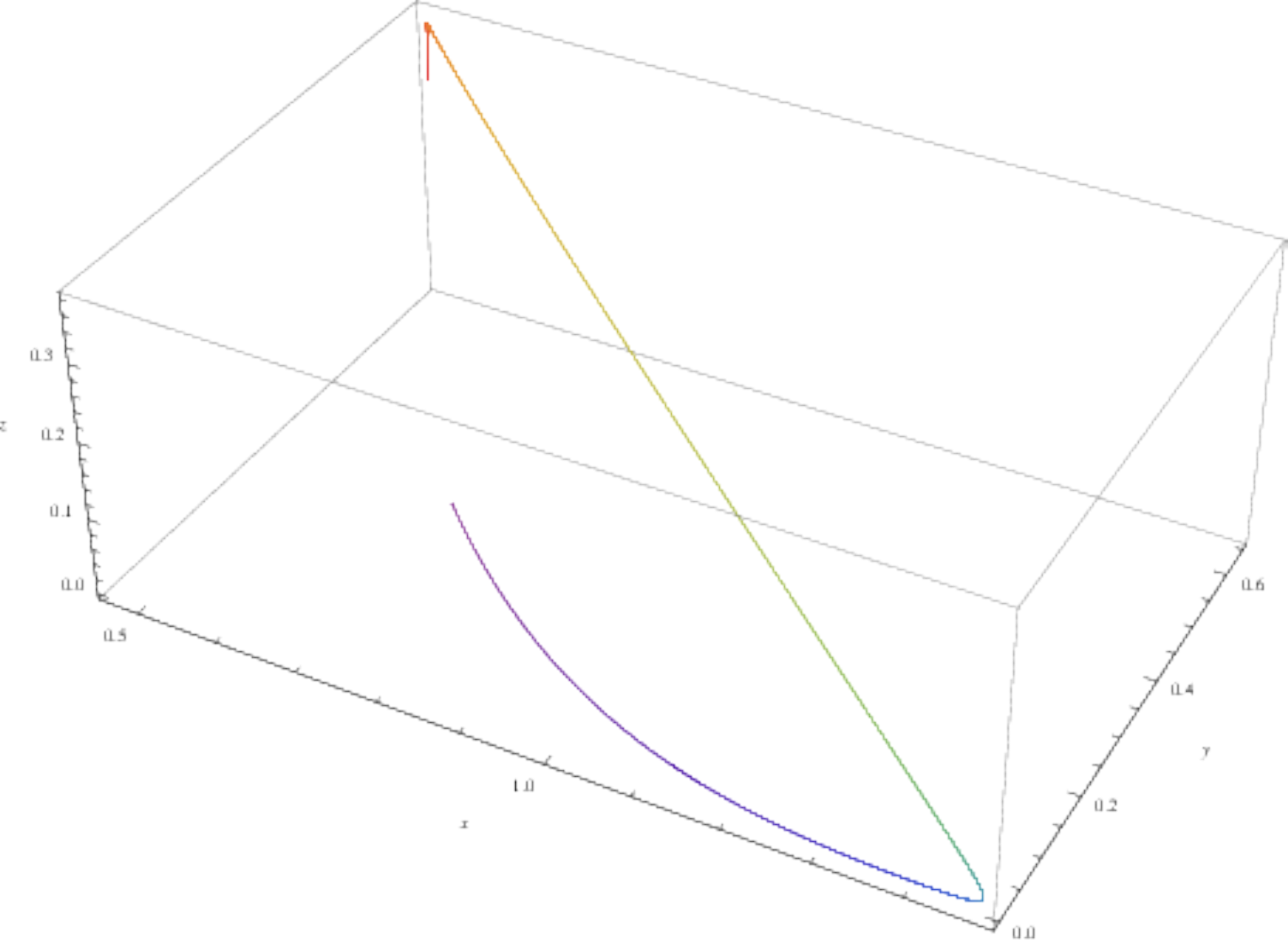}\\
   \end{tabular}
   \caption{ Model III:  Phase space for $\tilde{\omega}=(-1,-1.2,-1.5),\, b=0.5,\,\omega_r=\frac{1}{3}$}
  \label{fig6}
   \end{figure}

  $P_{31}$ is stable for $w_d>-1$. $P_{32}$ , $P_{34}$ are unstable.  $P_{33}$ 
  is systematically stable when $w_d<-1$, $b<\frac{1-3w_d}{9(1+\alpha)}$.

  We present here the dynamic of model {\it III} through the figures \ref{fig5} and \ref{fig6}. Here we note a gradual increase for the dark energy 
  whereas the energy densities of  the radiation and the matter  are tending to zero. These facts are compatible with the recent observational data
  showing that Universe is accelerated expansion because of the strong presence of dark energy in Universe. This analysis shows also that becomes 
  nonexistent because of the strong domination of dark energy. The phase diagram of interaction between dark  energy and the both matter and radiation 
  is plotted in figure \ref{fig6}. In parallel with the previous models,  this model is also one of those
   where the behavior  the energy density of dark energy is that of quintessence while the densities of radiation and matter are going to vanish
   during the
    the expansion.
 \subsection{Interacting Model - IV}
Let's search for new model generalized by the new following interaction terms: 
\begin{eqnarray}\label{11}
E_1&=&-3b\kappa^2 H^{-1}\tilde{\rho}\rho_r,\nonumber\\
E_2&=&3b\kappa^2 H^{-1}(\tilde{\rho}\rho_r-\rho_m\rho_r),\nonumber\\
E_3&=&3b\kappa^2 H^{-1}\rho_m\rho_r.
\end{eqnarray}

 The system in (\ref{mouss}) takes the form

\begin{eqnarray}
\frac{dx}{dN}&=&3x\Big(x+y+z+\tilde{\omega}x+\omega_r z\Big)-3x-3\tilde{\omega}x-9bxz,\nonumber\\
\frac{dy}{dN}&=&3y\Big(x+y+z+\tilde{\omega}x+\omega_r z\Big)-3y+9b(xz-yz),\\
\frac{dz}{dN}&=&3z\Big(x+y+z+\tilde{\omega}x+\omega_r
z\Big)-3z-3\omega_rz+9byz.\nonumber
\end{eqnarray}

 \begin{figure}[h]
   \centering
   \begin{tabular}{rl}
   \includegraphics[width=8cm, height=6cm]{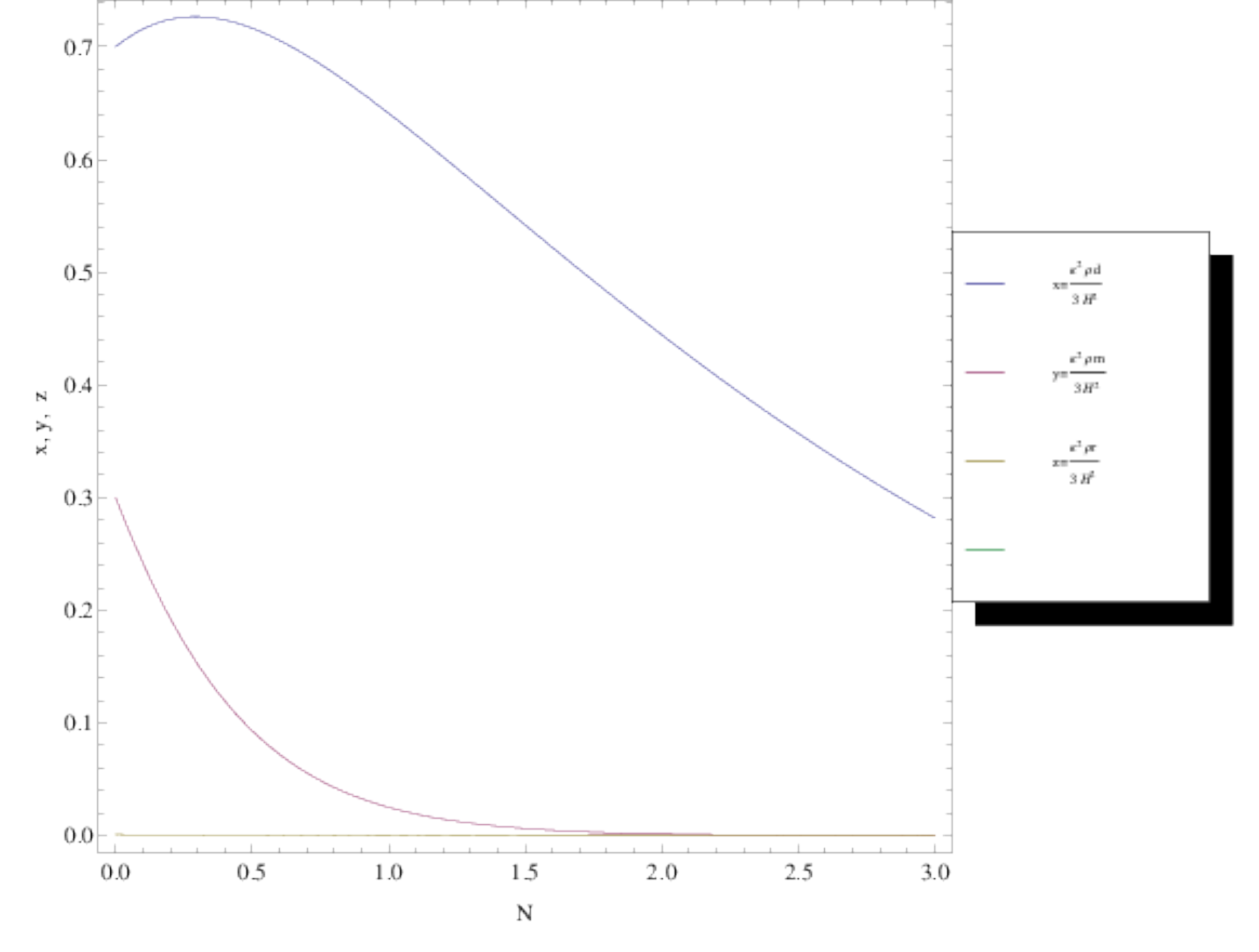}
   \end{tabular}
   \caption{ Model IV:   variation of $x,y,z$ as a function of the $N=\ln(a)$. The initial
   conditions chosen are $x(0)=0.7,y(0)=0.3,z(0)=0.01$,
   $\tilde{\omega}=-1.2,\,\,\omega_r=\frac{1}{3}$ and $b=0.5$.}
  \label{fig7}
   \end{figure}

 One obtains seven critical points

 \begin{tabular}{|c|l|p{4cm}|m{4cm}|b{4cm}|}
 \hline 
 Points & $\lambda_1$&$\lambda_2$ &$\lambda_3$ &  $(x_c,y_c,z_c)$
 \\\hline
 $ P_{41}$ &$-3$ & $ -4$& $-3(1+\tilde{\omega})$ &  $\ (0,\quad 0,\quad0)$
 \\\hline
 $ P_{42}$ & $3$ &$-1$&$ -3\tilde{\omega} $ & $\ (1,\quad 0,\quad0)$
 \\\hline
  $ P_{43}$   & $  3\tilde{\omega}$& $  3\tilde{\omega}-1 $ & $ 3\tilde{\omega}-1$ & $\ (0,\quad 1,\quad 0) $
 \\\hline
  $ P_{44}$   & $3(1+\tilde{\omega})$& $ -9b+1$ & $ -3\tilde{\omega}-9b+1  $ & $ \ (0,\quad0,\quad1)$ 
  \\\hline
  $ P_{45}$   & -& -& - & $ \ \Big(\frac{1}{3}\frac{\frac{4}{3}\tilde{\omega}}{b(1+\tilde{\omega})},\frac{4}{9b},
  -\frac{1}{3}\frac{1+\tilde{\omega}}{b}\Big)$ 
  \\\hline
  $ P_{46}$   & -& - & -& $ 
  \Big( \,{\frac {\tilde{\omega}+3\,b+3\,-1/3}{3b}}, \newline
  -\,{\frac { \left( -\frac{1}{3}+\tilde{\omega} \right)
  \left(\tilde{\omega}+3\,b+-\frac{1}{3} \right) }{3b \left(
 3\,b-\frac{1}{3} \right) } } ,\newline  
 {\frac
 {\tilde{\omega} \left( -\frac{1}{3}+\tilde{\omega} \right) }{3b \left( 3\,b-\frac{1}{3} \right) }}\Big)$ 
 \\\hline
  $ P_47$   & $-3\tilde{\omega}$& $ \frac {\sqrt {3}\sqrt {b 
 \left(-4/9+3\,b+\,b \right) }}{b }$ & 
 $ -{\frac {\sqrt {3}\sqrt {b \left(-4/9+3\,b+\,b \right) }}{b  }} $ & $ \  (0, \frac{4}{9b}, -\frac{1}{3b})$ 
  \\\hline
 \end{tabular}
   
 \begin{figure}[h]
   \centering
   \begin{tabular}{rl}
   \includegraphics[width=8cm, height=6cm]{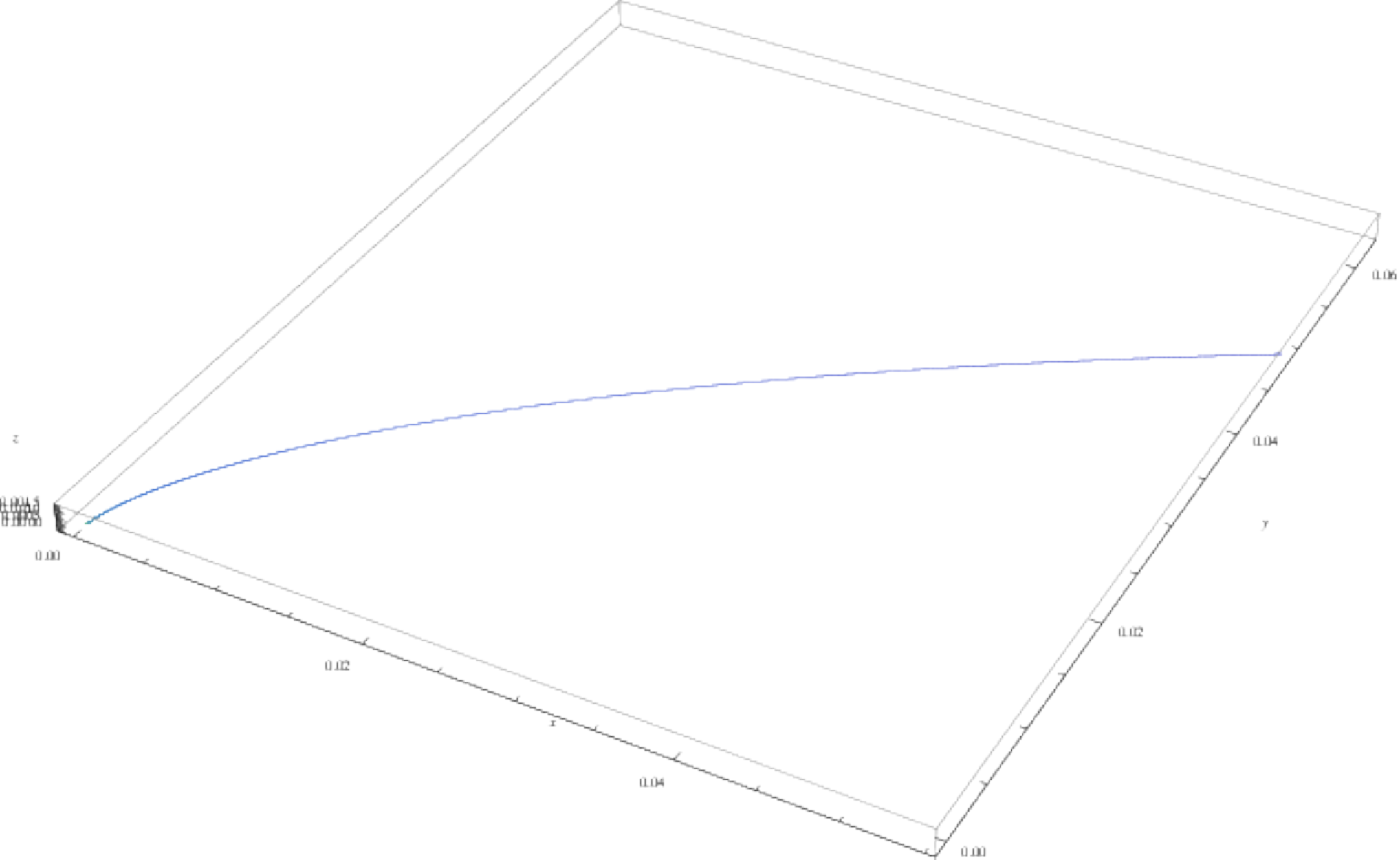}&
    \includegraphics[width=8cm, height=6cm]{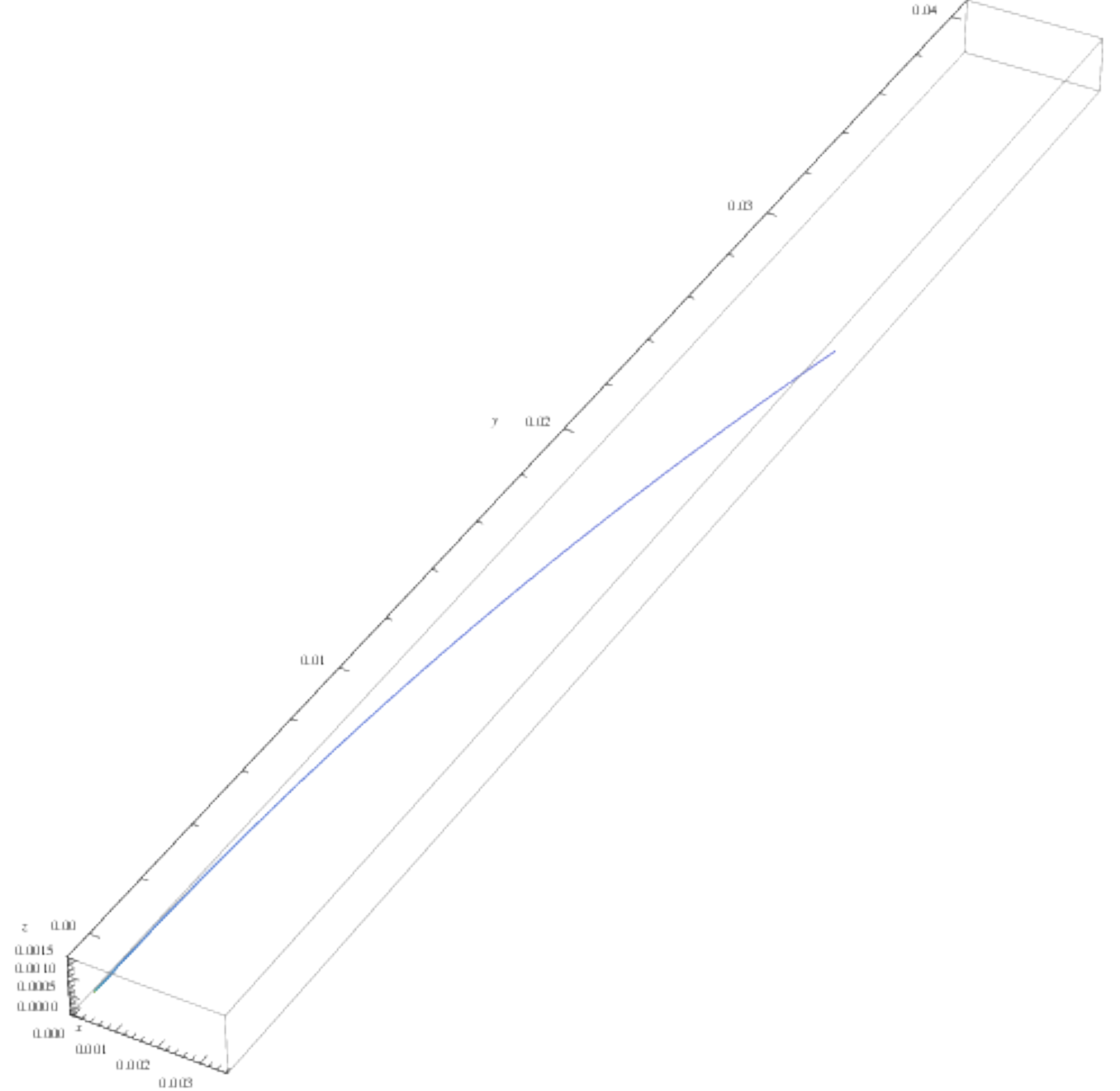}\\
  \end{tabular}
   \caption{ Model IV:  Phase space for $\tilde{\omega}=(-1,-1.2,-1.5),\, b=0.5,\,\omega_r=\frac{1}{3}$}
  \label{fig8}
   \end{figure}

We remark  for this model that $ P_{41}$ is stable for $\tilde{\omega}>-1$ ,  $ P_{42}$ is also stable for  $\tilde{\omega}<-1$ while $ P_{43}$ and  $ P_{47}$ are unstable. $ P_{44}$ is stable for $\tilde{\omega}<-1$ and $b>\frac{1}{9}$.  It is also possible to determine the stability of point $ P_{45}$. The stabilities of the  point  $P_{47}$  and also  $ P_{35}$ from the model III are not easy to be studied because the matrix of Jacobi in these cases is not diagonal  and the behaviors of its eigenvalues are not trivial.  This means that we can not theoretically know if these points are stable or not. Consequently these cases are analytically and numerically impossible to be studied. Therefore, the dynamic behaviors of the model IV behavior have been plotted in the figures \ref{fig7} and  \ref{fig8} which show that the density of 
 the dark energy quickly rise up to $N \sim 0,5$ and after decreases sharply (same behavior with quintessence) whereas the energy
 densities of radiation and  matter   decrease and tend to zero when  $N \sim 1,8$.

 \section{Stability of $  T+ \mathcal{T}^{\mathcal{N}} $ model}\label{sec5}
 
 This section is devoted to the study of the stability of model $ f(T, \mathcal{T})=  T+ \mathcal{T}^{\mathcal{N}} $ by using the power-law and the 
 de Sitter solutions.\par 
 We are interested here to the perturbation of both geometric parts and matter of the generalized equations of motion.  To do so, we have focused our 
 attention on the  Hubble parameter for geometric perturbation and energy density for ordinary primordial matter perturbation and we have followed the 
 same way as it is done in  \cite{antoniode, diegoalvaro}.
 
\begin{eqnarray} 
 H(t) = H_{b}(t)(1+\delta(t)),\quad   \rho(t)= \rho_{b}(t) (1+\delta_{m}(t))\label{7}.
\end{eqnarray}  
 $H_b(t)$ and $\rho_b(t)$ denote the Hubble parameter and the energy density of the ordinary matter of the background respectively.
 Taking into consideration   the interaction term, the continuity equation of the ordinary matter becomes the following differential equation

\begin{eqnarray}    
\dot{\rho_{b}}(t)+3H_{b}(t)\rho_{b}(t)(1+\omega+q)= 0\,, \label{8}
\end{eqnarray}        
whose resolution leads to: 
\begin{eqnarray}
\rho_{b}(t)=\rho_{0}\ e^{-3(1+\omega+q)\int {H_{b}(t)}dt},\label{9}
\end{eqnarray}
where $\rho_{0}$ is an integration constant. In order to study the linear perturbation about $ H(t)$ and 
 $ \rho(t)$, we develop  $\mathcal{T}^{\mathcal{N}} $ in a series of $\mathcal{T}_{b} = \rho_{b}(1-3\omega)$ as:
\begin{eqnarray}
f(\mathcal{T}) = f^{b}+f_{\mathcal{T}}^{b}(\mathcal{T}-\mathcal{T}_b) + O^{2}\label{10}, 
\end{eqnarray}
  The function $\mathcal{T}^{\mathcal{N}} $ and its derivatives are computed at  $\mathcal{T}=\mathcal{T}_{h}$. 
  According to the Einstein-Hilbert term, the strangeness here is the effect coming from $  \mathcal{T}^{\mathcal{N}} $. By putting (\ref{7}) into 
   (\ref{10}) in the the first generalized Friedmann equation; one gets 
\begin{equation}
3H^{2}= \rho-\frac{1}{2}\left(f+12 H^{2} f_{T}\right)+ 3f_{%
\mathcal{T}} \left(\frac{\rho + p}{3}\right),
\label{Friedmann1}
\end{equation}
which gives after simplification
\begin{eqnarray}
6H_{b}^{2}(t)\delta(t)= \big[\rho_{b}+\rho_{b}f_{\mathcal{T}}^{b}(\frac{3-\omega}{2})+
\rho_{b}^{2}(1-2\omega-3\omega^{2})f_{\mathcal{T}\mathcal{T}}^{b}\big]\delta_{m}(t).
\label{11}
\end{eqnarray}
 Considering that the ordinary matter is essentially the dust, we obtain the simple expression
\begin{eqnarray}
6H_{b}^{2}(t)\delta(t)= \big[\rho_{b}+3\rho_{b}f_{\mathcal{T}}^{b}+
2\rho_{b}^{2}f_{\mathcal{T}\mathcal{T}}^{b}\big]\delta_{m}(t),\label{12}.
\end{eqnarray}
For matter pertubation function, we have the following differential equation 
\begin{eqnarray}
\dot{\delta_{m}}(t)+3(1+\omega+q)H_{b}(t)\delta(t) = 0. \label{13}
\end{eqnarray}
 Eliminating $\delta(t)$ between (\ref{11}) and (\ref{13}), we obtain also the differential equation 
\begin{eqnarray}
2H_{b}\dot{\delta}_{m}(t)+(1+\omega+q)
\big[\rho_{b}+\rho_{b}f_{\mathcal{T}}^{b}\left(\frac{3-\omega}{2}\right)+
\rho_{b}^{2}(1-2\omega-3\omega^{2})f_{\mathcal{T}\mathcal{T}}^{b}\big]\delta_{m}(t)=0.\label{14}
\end{eqnarray} 
 The direct resolution of this differential equation gives
\begin{eqnarray}
\delta_{m}(t)= C_{0}\exp\left\{-\left(\frac{1+\omega+q}{2}\right)\int 
\frac{\rho_{b}}{H_{b}}\left[1+f_{\mathcal{T}}^{b}\left(\frac{3-\omega}{2}\right)+\rho_{b}
(1-2\omega-3\omega^{2})f_{\mathcal{T}\mathcal{T}}^{b}\right]dt \right\} ,
\label{15} 
\end{eqnarray}
where $C_{0}$ is an integration constant. From Eq.~(\ref{13}) one can  extract 
\begin{eqnarray}
\delta(t) = \frac{C_{0} C_{\mathcal{T}}}{6H_{b}}\exp\left\{ -\left(\frac{1+\omega+q}{2}\right)\int C_{\mathcal{T}}dt \right\}  ,
\label{16}
\end{eqnarray}
with
\begin{eqnarray}
C_{\mathcal{T}}=\frac{\rho_{b}}{H_{b}}\left[1+f_{\mathcal{T}}^{b}
\left(\frac{3-\omega}{2}\right)+\rho_{b}\left(1-2\omega-3\omega^{2}\right)f_{\mathcal{T}\mathcal{T}}^{b}\right].
\end{eqnarray}

 \subsection{Stability of de Sitter solutions}\label{sec5.1}
 In this case, the Hubble parameter is written as  
\begin{eqnarray}
H_{b}(t) = H_{0} \rightarrow a(t) = a_{0}e^{H_{0}t}.
\end{eqnarray}
The expression (\ref{9}) becomes,
\begin{eqnarray}
\rho_{b}(t)=\rho_{0} e^{-3(1+\omega+q)H_{0}t}.
\label{17}
\end{eqnarray}
 From relation $ d{\rho_{b}}=-3(1+\omega+q)H_{0}\rho_{b}dt$ and within an elementary transformation, we get 
\begin{eqnarray}
\int C_{\mathcal{T}}dt &=&-\frac{1}{3H_{0}(1+\omega+q)}\int\frac{1}{\rho_{b}}C_{\mathcal{T}}d{\rho_{b}} \nonumber\\
&=&-\frac{1}{3H_{0}^{2}(1+q+\omega)}\Bigg\{ \rho_{b}
+Q\; \frac{(3-\omega)}{2}    \rho_{b}^{\mathcal{N}}(1-3\omega)^{\mathcal{N}-1} 
\nonumber\\
&+& Q\;(1-2\omega-3\omega^{2}) (\mathcal{N}-1)(1-3\omega)^{\mathcal{N}-2}\rho_{b}^{\mathcal{N}} \Bigg\}.
\end{eqnarray}
By replacing this expression in  \ref{15}, we  obtains 
\begin{eqnarray}
\delta_{m}(t)= C_{0}\exp \left\{ \frac{1}{6H_{0}^{2}}\left[\rho_{b}+Q\; \frac{(3-\omega)}{2}    \rho_{b}^{\mathcal{N}}(1-3\omega)^{\mathcal{N}-1}
+ Q\;(1-2\omega-3\omega^{2}) (\mathcal{\mathcal{N}}-1)(1-3\omega)^{\mathcal{N}-2}\rho_{b}^{\mathcal{N}}\right]\right\}.
\label{18}
\end{eqnarray}
 Therefore the perturbation function about the geometry can be obtained and  given by
\begin{eqnarray}
\delta(t) &=& \frac{C_{0} C_{\mathcal{T}}}{6H_{0}}\exp \left\{ \frac{1}{6H_{0}^{2}}\left[\rho_{b}+Q\; \frac{(3-\omega)}{2} 
\rho_{b}^{\mathcal{N}}(1-3\omega)^{\mathcal{N}-1} +  
Q\;(1-2\omega-3\omega^{2}) (\mathcal{N}-1)(1-3\omega)^{\mathcal{N}-2}\rho_{b}^{\mathcal{N}}\right]\right\},
\label{19}
\end{eqnarray}
with
\begin{eqnarray}
 C_{\mathcal{T}}= \frac{1}{H_{0}}\left\{\rho_{b}+Q\;\mathcal{N} \frac{(3-\omega)}{2}    \rho_{b}^{\mathcal{N}}(1-3\omega)^{\mathcal{N}-1} + Q\;\mathcal{N}  (\mathcal{N}-1)(1-2\omega-3\omega^{2})(1-3\omega)^{\mathcal{N}-2}\rho_{b}^{\mathcal{N}}\right\}
\end{eqnarray}
 and

\begin{eqnarray}
 f(\mathcal{T}) = Q\; \mathcal{T}^{\mathcal{N}}. 
\end{eqnarray}

 with  $Q$ the one defined in (\ref{cvalue}) and $\mathcal{N} =-(1-3\omega)/((1+\omega))$.
 For some suitable values of the input parameters consistent with cosmological observational data, we plot the curve
 characterizing the behavior of the perturbation function at the left side in Fig \ref{fig2}. We see that as the universe expands, i.e.,
 increasing $Z$, the matter and geometric perturbations functions, $\delta_m$ and $\delta$ respectively, goes towards positive values
 more less than $0.1$ when the time evolves.
 
  \subsection{Stability of Power-Law solutions}\label{sec5.2}
\begin{figure}[h]
 \centering
 \begin{tabular}{rl}
 \includegraphics[height=5cm,width=8.5cm]{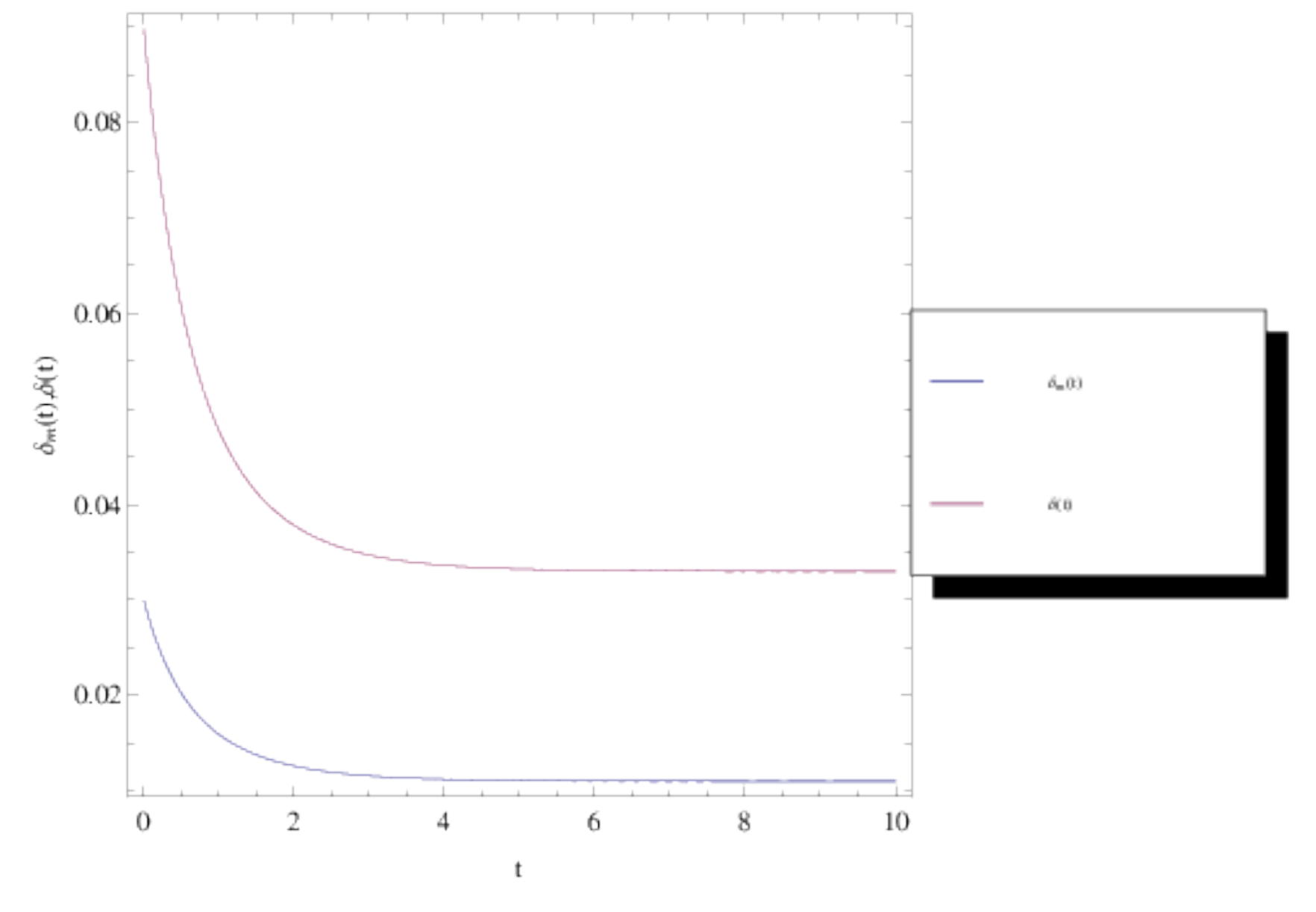} \label{e3}&
 \includegraphics[height=5cm,width=9cm]{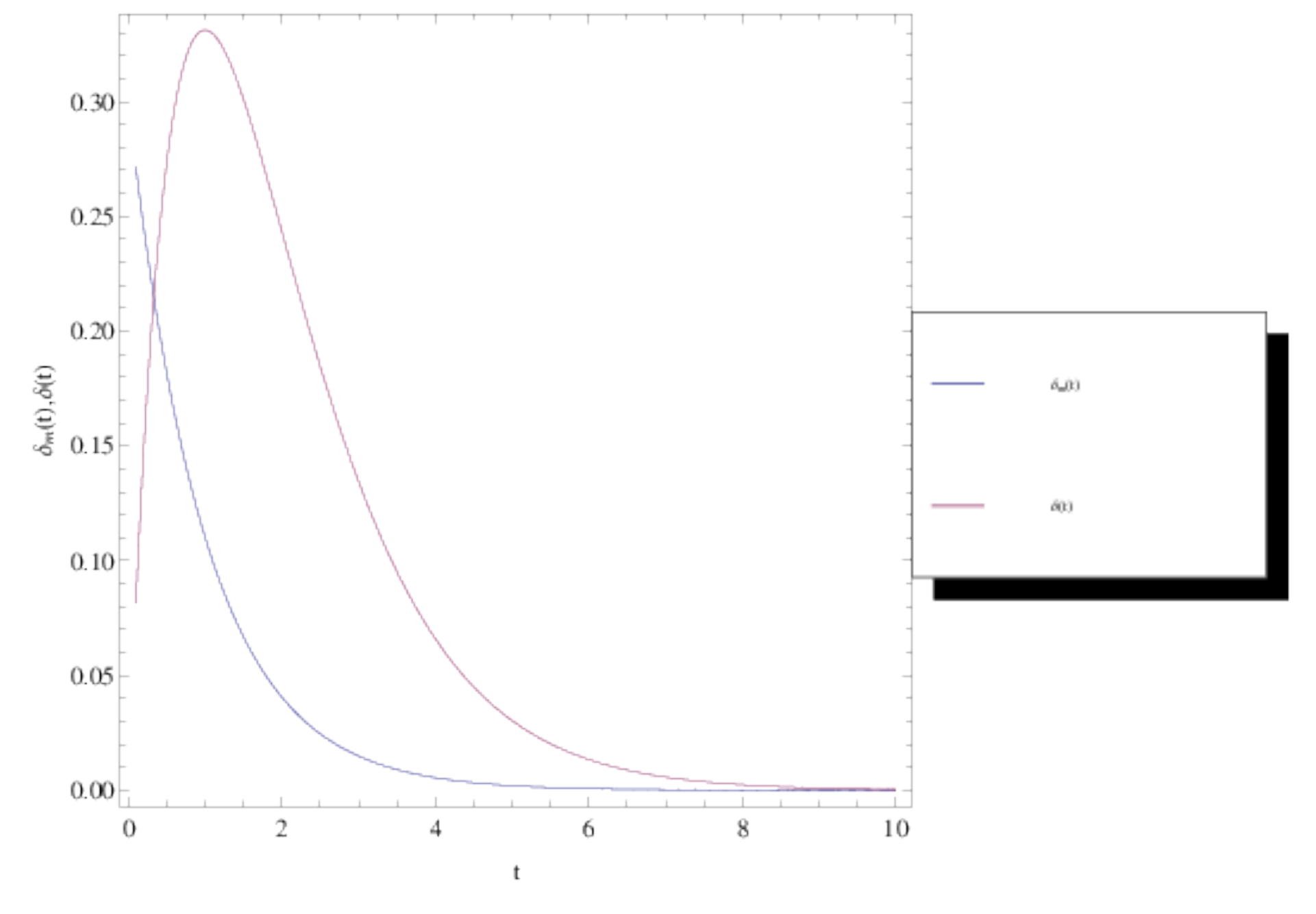} \label{e4}
 \end{tabular}
 \caption{ The graph at the left side of the figure presents the evolution of the perturbation functions $\delta_m$  
 and $\delta$  within the de Sitter solutions, while the one at the right side shows the evolution of the perturbation
 functions within the power-law solutions. 
 The graph are plotted for $n=2/3$, $\Lambda=1.7\times 10^{-121}$, $\rho_0=0.1\times 10^{-121}$, 
 $\omega=0$ and $C_1=1$ .}
 \label{fig21}
 \end{figure}

Here, the scale factor is written as 
\begin{eqnarray}
 a(t)\propto t^{n} \quad \rightarrow   H_{b}(t) = \frac{n}{t}\,,
\end{eqnarray} 
  and the ordinary energy density (\ref{9}) becomes
\begin{eqnarray}
\rho_{b}=\rho_{0}t^{-3n(1+\omega+q)}\label{20}
\end{eqnarray}
 By making the substitution of  $\rho_{b}$ in  (\ref{14}), one gets after resolution, the following expression
\begin{eqnarray}
\delta_{m}(t)= C_{1}\exp\left\{ -A\left(\frac{A_{1}}{2+B}t^{2+B} +\frac{A_{2}}{2+\mathcal{N} B}t^{2+\mathcal{N} B}\right)\right\},
\end{eqnarray}
 with $C_{1}$ an integration constant, and
\begin{eqnarray} 
A = \frac{(1+\omega+q)}{2n}, \quad A_{1}= \rho_{0},\quad B = -3n(1+\omega+q)\nonumber\\ A_{2}=Q\;\mathcal{N} \rho_{0}^{\mathcal{N}}\left\{ \frac{(18\omega^{3}+9\omega^{2}-14\omega+3)}
{2(1-3\omega)^{2-\mathcal{N}}}+\frac{\mathcal{N}}{{(1-3\omega)}^{(1-\mathcal{N})}}\right\}.
\end{eqnarray} 
  The use of  the relation (\ref{13} leads to
\begin{eqnarray} 
\delta(t)=\frac{C_{1}}{6n^{2}}\left( A_{1}t^{2+\mathcal{N}}+A_{2} t^{2+B \mathcal{N}}\right) \exp\left\{ -A\left(\frac{A_{1}}{2+B}t^{2+B} +\frac{A_{2}}{2+\mathcal{N} B}t^{2+\mathcal{N} B}\right)\right\}.
 \end{eqnarray}
 As we have done in the previous section, we present here the evolution of the perturbation functions in  Fig \ref{fig21} for suitable values of input parameters.

\newpage

\section{Conclusion}\label{sec6}
 We undertook in this work cosmological analysis about a model in the framework of the so-called $f(T, \mathcal{T})$ theory. In
 order to obtain a viable $f(T, \mathcal{T})$ model, we first impose the covariant conservation of the energy-momentum, from which, 
 we get a model of the type $T+f(\mathcal{T})$, being a sort of trace depending function correction to the TT. 
 The  obtained model includes parameters depending on the cosmological constant $\Lambda$ and the parameter $\omega$ of the
 ordinary equation of state. These parameters play a main role in the whole study developed in this manuscript. By the way, 
 we study the dynamics of the cosmological system, analyzing the stability about the critical points.  We solve the equations and it appears that for some specific expressions of the interaction term one can obtain attractor  solutions. We  numerically integrate the equations and show that the evolution of  the dark energy density mimics three diffract behaviors: phantom,  quintessence and cosmological constant in some interactive forms.  We argue that this interaction is purely phenomenological and is consistent with the observational data. Our result shows  that for both de Sitter and power-law solutions, the perturbations functions converge traducing the stability of the model. \par
 Moreover, the stability of the model is checked within the de Sitter and power-law solutions by performing linear perturbation 
 about the physical critical points. We see that for the both considered  solutions, the model presents stability through the
 convergence of the geometric and matter perturbation functions $\delta$ and $\delta_m$.

\section*{Acknowledgments}
 The authors  thank IMSP for hospitality during the elaboration of this work.

\end{document}